\documentclass[11pt,a4paper]{article}%\documentclass[bibliography=totocnumbered,listof=totocnumbered,openany,12pt]{scrbook}
\usepackage[makeroom]{cancel}
\usepackage{bm}
\usepackage{amsmath}
\usepackage{amsfonts}
\usepackage{amssymb}
\usepackage{graphicx}
\usepackage{nicefrac}
\usepackage{authblk}
\usepackage{cite}
\usepackage[left=2.54cm,top=2.54cm,right=2.54cm,bottom=2.54cm,nohead]{geometry}
\DeclareGraphicsExtensions{.png,.jpg,.pdf}
\usepackage{epstopdf}
\usepackage{float}
\usepackage{fouriernc} 
\usepackage{enumitem}
\usepackage[utf8]{inputenc}
\usepackage[english]{babel}
\usepackage{mathtools}
\usepackage{amscd}
\usepackage{mathrsfs}
\usepackage{amsthm}
\usepackage{siunitx}%For SI units
\usepackage{caption}
\usepackage{subcaption}
\setlist{nolistsep,leftmargin=*}
\usepackage{hyphenat}
\usepackage{mhchem} 
\usepackage{cleveref}
\usepackage{booktabs} %% nicer tables
\usepackage{multirow,rotating,adjustbox} %% for big table multirow vertical text

\title{Theory of ion and water transport \\ in reverse osmosis membranes}

\makeatletter
\renewcommand\AB@authnote[1]{\textsuperscript{\normalfont#1}}

\makeatother
\author[]{Y.S. Oren}
\author[]{P.M. Biesheuvel}
\affil[]{Wetsus, European Centre of Excellence for Sustainable Water Technology \\ Oostergoweg 9, 8911 MA Leeuwarden, The Netherlands}
\date{} %remove date

\newcommand{\s}[1]{\mathrm{_{#1}}}

\begin{document}

\maketitle

\begin{abstract}

We present theory for ion and water transport through reverse osmosis membranes based on a Maxwell-Stefan framework combined with hydrodynamic theory for the reduced motion of particles in thin pores. We include all driving forces and frictions both on the fluid (water), and on the ions, including ion-fluid friction as well as ion-wall friction. By including the acid-base character of the carbonic acid system, the boric acid system, H$_3$O$^+$/OH$^-$, and the membrane charge, we locally determine pH and thus the effective charge of the membrane as well as the dissociation degree of boric acid. We present calculation results for a ``dead end'' experiment with fixed feed concentration, where effluent composition is a self-consistent function of fluxes through the membrane. Comparison with experimental results from literature for fluid flow vs. pressure, and for salt and boron rejection, shows that theory agrees well with data. Our model is based on realistic assumptions for the effective sizes of the ions and for the diameter of the RO membrane pore in the polyamide toplayer ($\sim 0.75$ nm).
 
\end{abstract}

\section{Introduction}
In a world where potable water is a scarce natural resource, a need develops to produce it by artificial means~\cite{UN-Water2006}. This trend is expected to increase the consumption of desalinated water dramatically over the coming years~\cite{Burbano2007}. %Therefore, more and more countries are turning to desalination solutions as a means to produce drinking water. 
Reverse osmosis (RO) has become an increasingly popular solution in many countries to provide potable water. % for their citizens. %Population growth and climate changes have led to an increase in fresh water demand worldwide .
About 98\% of Earth's water is seawater (SW) and brackish water (BW), making these attractive sources.% for potable water.  

The principle of RO is to apply high pressure on an aqueous solution which is in contact with a membrane that allows passage of water, but not of salts. If the pressure is high enough to overcome the osmotic pressure of the solution, water will pass the membrane, resulting in a low-in-salts permeate stream. The possibility of desalinating water by the RO principle was already researched in the 1850s \cite{Baker2004} but it was not until the 1960s, when cellulose acetate (CA) membranes were introduced, that RO gained real industrial potential for the production of desalinated water. Still, there were many issues to be solved in order to make RO commercially competitive, and ultimately the preferred technology. One of the major issues at that time was the high energy consumption required to operate an RO desalination plant. Since the osmotic pressure of seawater is about thirty bars, the pressure that has to be applied to overcome the osmotic pressure is ``lost'' with the brine. In the 1980s energy recovery devices for RO were introduced, to recover the potential energy in the brine. %From Francis and Pelton turbines to turbochargers and piston exchangers ERDs are major contributors to the reduction in energy consumption of RO desalination plants. 
Together with the introduction of thin-film composite (TFC) polyamide (PA) membranes that increased permeate flux and improved rejection \cite{Larson1981}, RO energy consumption decreased dramatically over the years, to values that nowadays are only two or three times thermodynamic minimum, making RO currently the most energy-efficient desalination method \cite{Elimelech2011}. Besides production of water for human consumption and agriculture, RO technology has many different applications, such as in food processing, and use in the pharmaceutical, textile and paper industries \cite{Malaeb2011}.

%\indent Several industrial-mature desalination technologies are available. In general they can be classified into two groups – thermal methods such as multi-stage flash (MSF) and multiple effect distillation (MED), and membranal methods such as electrodialysis (ED) and RO. Nowdays, RO is a vastly used method for desalination of SW and BW worldwide, estimated to constitute about 60\,\% of the global desalination market (Fig. \ref{fig:GlobalTech}).\\
%\indent To demonstrate the fast growth of RO in the market, we can look at MSF as an example. Starting already in the 1960s, this technology was almost solely responsible to the global seawater desalination at that time. In 2006 it was estimated to produce 40\'\% of the global desalination capacity \cite{Zhou2005} and by 2012 it was estimated to hold only about 20\,\% (Fig \ref{fig:GlobalTech}).

In contrast to the immense progress in RO development over the past decades,  the actual transport and separation mechanisms are not yet fully understood \cite{Palmeri2006, Bason2011, Nir2015b}. In the attempt to describe the physico-chemical processes that govern the transport and separation process in RO membranes many mathematical models were developed throughout the years. Among those models are solution-diffusion models, which assume constant pressure and concentration gradients as the only driving force in the process, pore-flow models based on the pressure difference as driving force, and molecular dynamics simulations that address the mechanism on the molecular level, thereby taking the membrane composition also into account. In addition, irreversible thermodynamics models use a phenomenological approach to describe membrane transport~\cite{Malaeb2011,Wijmans1995,Shen2016}.

In our present work, we aim to describe the mechanisms by which ions are transported in RO membranes using a Maxwell-Stefan approach combined with information from hydrodynamic theory of particles in thin pores. We restrict our model to the polyamide toplayer of a thin-film composite (TFC) membrane and the concentration polarization layer located in front of the membrane. We will restrict our calculations to the desalination of artificial seawater. We study the effect of applied pressure and membrane charge on the composition of the permeate, such as pH and rejection of salt, and compare with literature values.

\section{Theoretical background}

\subsection{RO membranes}

RO membranes have come a long way since the start of development of RO membranes for seawater desalination in the 1960s. The first membranes were made of cellulose acetate (CA) with the most significant development achieved by Loeb and Sourirajan~\cite{Loeb}. Their asymmetric membrane, a 200 nm CA film on top of a porous support, laid the foundation for industrial-scale seawater desalination by RO \cite{Lee2011a}. Although the CA membranes allowed for at  least an order of magnitude higher water flux than common membranes available at the time, these membranes were not durable when exposed to pH changes, chlorine and microbial contamination. Since then, many types and forms of RO membranes have been developed. Today the most common RO membrane used in commercial installations around the world is a TFC membrane with a polyamide (PA) toplayer. This material was introduced in the 1970s \cite{Li2010} and provides a better performance than CA membranes in terms of flux, salt rejection, pH tolerance and operational temperature range \cite{Lenntech,Li2010}. These membranes consist of a very thin layer of aromatic PA (0.05--0.15 $\mu$m thickness) on top of a microporous support (pores 2 nm \cite{Sefara2007},  40 $\mu$m thickness) and a fabric layer (120--150 $\mu$m thickness) which supplies mechanical strength \cite{Petersen1993,Lee2011a,Ghosh2008}. The PA layer is created through on-surface polymerization of diamine and tricarboxyl monomers on the supporting structure leading to the polyamide structure. The PA layer is a (highly) crosslinked polymer, and thus one of the monomers contains at least three functional groups – two for propagation and one for crosslinking. The membrane layer can subsequently be coated with another polymer layer (e.g. polyvinyl alcohol, polyethyleneimine), for instance to supply additional protection from fouling \cite{Petersen1993,Lee2011,Li2010}, but this layer can also cause flux reduction \cite{Li2010}.

%Water is a polar molecule with a permanent dipole. If we take for example an ion in water, we can expect that the water molecules will arrange themselves around it, in accordance to the charge it carries. These ion-dipole interactions are dependent on the distance between the solute and solvent molecules, as well as the charge density of the ion \cite{Atkins2009}. If we consider two spherical ions that carry the same total charge, the ion that has the smaller ionic radius will have more charge per unit of surface area, hence will have higher charge density, and will have stronger interaction with the water molecules. In general, because cations "lose" and anions "gain" electrons, anions' ionic radii exceed their atomic radii and vice versa for cations. A solvation shell can also form around molecules that don't carry charge because of their permanent, induced or self-induced dipole (polarizability). Overall, neutral molecules possess weaker solvation shells \cite{Pagni2006}.

Essential in the PA toplayer is obviously the presence of pores. In RO, the pores in the membrane refer to the percolated free volume that is present between the polymer chains in the PA toplayer. The number of pores and the pore size distribution depend on the polymerization process~\cite{Shen2016}. Relative to nanofiltration membranes, RO membranes have a more narrow size distibution, i.e., a more uniform pore size~\cite{Kosutic2006}, typically in the range of an average diameter between 0.66 and 0.78 nm~\cite{Kosutic2006,Kezia2013,Yoon2005}. Kim \textit{et al.} \cite{Kim2005} distinguished between two types of pores. The smaller type, ``network pores'', are defined as the gaps between polymer branches, while ``aggregate pores'' relate to the spaces between polymer aggregates. Network pores have a smaller size ($\sim$ 0.4--0.5 nm) and constitute about 70\% of the pore volume, while aggregate pores ($\sim$ 0.7--0.9 nm) account for the remainder~\cite{Coronell2008,Kim2005}.

The charge of TFC RO membranes is a consequence of the polymerization process. Carboxylic and amine functional groups that did not participate in the crosslinking process or in chain elongation reactions, remain free and can be protonated or deprotonated depending on pH~\cite{Coronell2008}. Protonated amine groups result in positive charge and deprotonated carboxyl groups in negative charge. This means that the membrane can have a positive or negative charge which can be described as a function of pH, as will be discussed in more detail in section \ref{subsection:Mem_charge}.

\subsection{Seawater characteristics}\label{seawater_characteristics}

Seawater is the largest water body available on earth -- a fact that makes it an attractive, inexhaustible source for drinking water. %The composition of seawater is determined by the inlet source (rivers) and by exposure to the atmosphere. 
Seawater is characterized by high salinity (``total dissolved solids'' (TDS) > 35 g/L) with sodium and chloride being the major ions. pH value of seawater is slightly basic and typically is around pH 8.0. With the development and advances in RO technology throughout the years, the usage of seawater as a source of potable water rises, at present estimated at 60\% of the global intake for RO desalination \cite{Bennett2013a}. Table \ref{Table:seawater} presents an example of a typical seawater composition.

\begin{table}[H]
\centering
\caption{Composition of mediterranean seawater for most common ions\cite{Fritzmann2007,Company}.}
\label{Table:seawater}
\begin{tabular}{lclll}
\toprule
Species & Concentration (mg/L)     &  &  &  \\
\midrule
Na$^+$   & 12,500 &  &  &  \\
Mg$^{2+}$  & 1,450  &  &  &  \\
Ca$^{2+}$  & 450    &  &  &  \\
K$^+$    & 450    &  &  &  \\
Cl$^-$   & 22,100 &  &  &  \\
SO$_4^{2-}$ & 3,410  &  &  &  \\
HCO$_3^-$ & 160    &  &  &  \\
B & 4-5      &  &  &  \\
pH    & 8.1      &  &  &  \\
\bottomrule 
\end{tabular}
\end{table}

\newpage

Several acid-base reactions occur in (sea)water, which must be included in the theory. In all cases we will assume that the reaction is infinitely fast, and thus, locally, the ions participating in the reaction are at chemical equilibrium with one another. First of all we must consider the auto-protolysis reaction of water, in which a proton is transferred from one molecule to the other by 

\smallskip

{\ce{2H2O <=> H3O+ + OH-} , pK$\s{w}=14$ @ 25 $^\circ$C, pure water}

\smallskip

\noindent The value of pK$\s{w}$ for pure water at 25 $^\circ$C of pK$\s{w}\sim$14, drops to lower values for more saline solutions~\cite{Dyrssen1973,Dickson1979}. 

Boron is an element present in natural water systems. Seawater contains about 5 mg/L of elementary boron~\cite{Dyrssen1973,Zeebe2001}, which is predominantly in one of two forms: boric acid $\ce{B(OH)3}$ and the borate ion $\ce{B(OH)4-}$. For boron concentrations higher than 22 mg/L, other species, mainly cyclic forms, may be present as well, depending on pH \cite{Tu2010}. The distribution between boric acid and borate is given by

\smallskip

\ce{B(OH)3 + 2H2O <=> H3O+ + B(OH)4-} , pK$=9.23$ @ 25 $^\circ$C, pure water

\smallskip

\noindent It was shown in ref.~\cite{Tu2010} that salinity affects pK, bringing it down to pK$=8.60$ for 40 g/L TDS. 

The degree of rejection of boron in seawater RO is of considerable concern. In low concentrations it has been shown that boron is beneficial to human health such as for cell metabolism, bone density and immune response \cite{Devirian,Mastromatteo1994}, as well as being essential to proper growth of plants. It participates in cell distribution, growth and metabolism, respiration regulation and photosynthesis. Documented cases of high boron intake occasionally reported abdominal pain, vomiting and diarrhea. In some cases fever, rash and muscle cramps were documented \cite{Litovitz1988,Linden1986}. Based on these and other findings regarding boron toxicity, and because of the incomplete understanding of the influence of boron on human health, the World Health Organization (WHO) originally recommended a guideline of a maximum concentration of 0.5 mg/L total boron in drinking water \cite{WHO2008}. Since 2009 the guideline recommended by the WHO is set at 2.4 mg/L~\cite{Edition2011} (2.0 mg/L according to ref.~\cite{Frisbie2015}). According to WHO, this value can be challenging to achieve, depending on the water source and the used treatment technology~\cite{Edition2011}. Though this limit seems sufficiently low for safe human consumption of potable water, this concentration is too high when the water is used for irrigation purposes, and lower boron concentrations must be reached, depending on crop type \cite{Ayers1985}. 

In desalination by RO, it is argued that boron in the acid (hydrated) form passes through the RO membrane quite freely while the borate ion is rejected. The borate ion is weakly hydrated and thus relatively small~\cite{Corti1980} and we can assume the same holds for the boric acid molecule. Since boron rejection depends on the distribution between the uncharged and charged form, determined by the equilibrium constant, pH and temperature are expected to influence boron rejection. In addition, the membrane itself can enhance borate ion rejection when it is negatively charged. To facilitate rejection, the pH of the feed has to be around 8.0 in order to achieve about 88--93\% rejection \cite{Guler2015}. %However, increasing pH causes precipitation (scaling) of poorly soluble compounds such as \ce{CaCO3} and \ce{MgCO3}. 
To further reduce boron concentration in the effluent, RO plants must use an additional RO step, boron selective resins and/or pH adjustments, to meet regulations on boron concentration~\cite{Hilal2011}.

The carbonate system functions as a buffer system in natural water and is the main contributor to the ability of seawater to buffer pH changes. The seawater carbonate system is in equilibrium with the \ce{CO2} in the air, which in its dissolved form becomes \ce{H2CO3} (carbonic acid). The equlibria in water are given by

\smallskip
{
\noindent \ce{CO2(g) <=> CO2(aq)}\\
\ce{CO2(aq) + H2O(l) <=> H2CO3(aq)}\\
\ce{H2CO3(aq) <=> H+ + HCO3-} , pK$_{\ce{H2CO3}}=6.33$ @ 25 $^\circ$C, pure water \\
\ce{HCO3- <=> H+ + CO3^2-{}} , pK$_{\ce{HCO3-}}=10.33$ @ 25 $^\circ$C, pure water}

\smallskip

Note that also for the carbonate system it was shown that pK is a function of salinity, resulting in pK\textsubscript{\ce{H2CO3}}$=5.98$ and pK\textsubscript{\ce{HCO3-}}$=9.16$ for a 30 g/L salt solution at 25 $^\circ$C \cite{Millero1979,Dickson1987}. In RO, it is known that the permeate is characterized by trace amounts of \ce{HCO3-} and \ce{CO3^2-}. The rejection of carbonate by RO membranes therefore increases carbonate concentrations in the brine, which can lead to precipitation of components such as \ce{CaCO3} and \ce{MgCO3} on the membrane surface, a phenomenon known as ``scaling.'' %is important because ions such as \ce{Mg^2+} and \ce{Ca^2+} can create insoluble compounds with \ce{CO3^2-} (scaling). %The occurrence of scaling, or precipitation, is dependent on factors such as pH, concentration, temperature and pressure. 

In our work, the above equilibria are combined with the relevant transport equations which are set up for each ion individually. The resulting coupled model of rejection of weak acid systems has an inherent complexity, which lies in the fact that in addition to transport, species can form and deplete due to the various chemical equilibria.

\subsection{Concentration polarization}

\begin{figure}[H]
\centering
\includegraphics[width=0.8\textwidth]{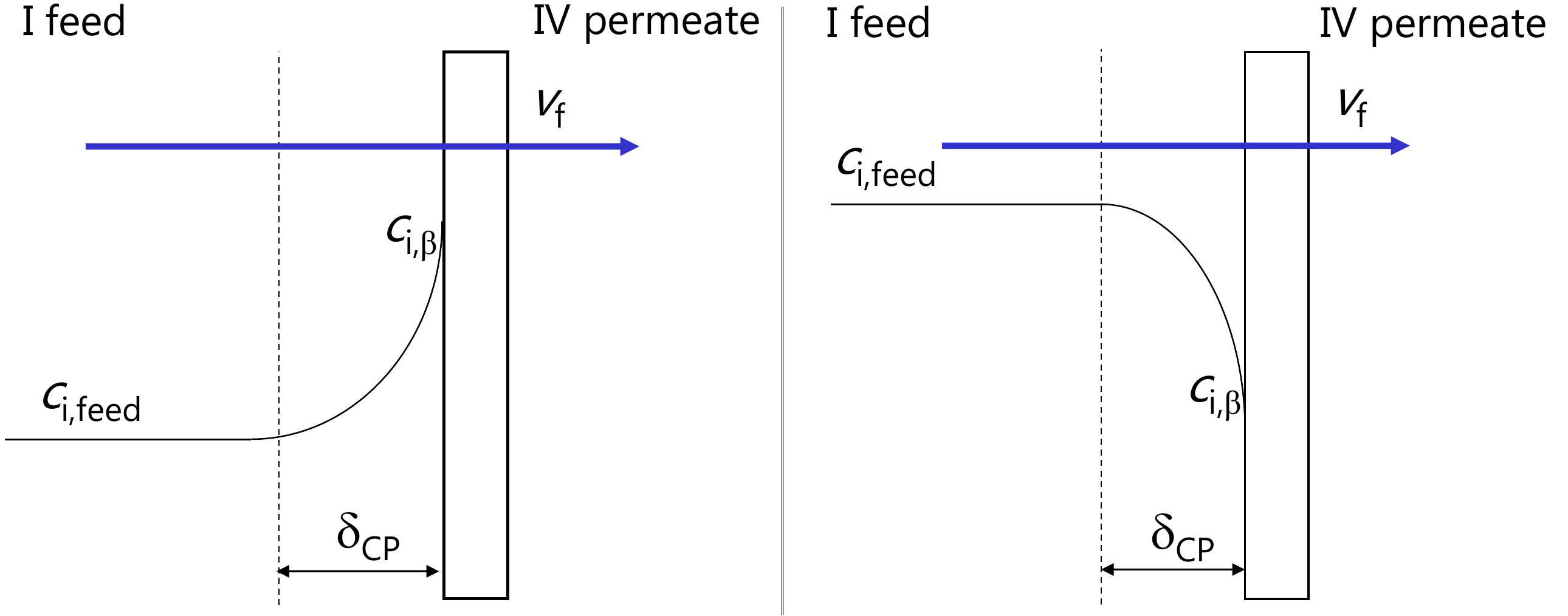}
\caption{Sketch of concentration polarization in reverse osmosis for the case where (a) a component is enriched at the membrane surface and (b) when a component is depleted. Situation (a) is the typical situation in RO when ionic species are rejected by the membrane, but for some species the opposite behavior is possible.} \label{fig:CP}
\end{figure}

Every membrane separation process promotes the passage of some species over others. This selective permeation leads to gradients of concentrations in the unmixed boundary layer in front of the membrane. For RO, if the species is rejected by the membrane it will accumulate on the surface resulting in a higher concentration (Fig. \ref{fig:CP}a) and if the membrane favors its passage, the ion will deplete (Fig. \ref{fig:CP}b). In RO, typically the salt concentration on the membrane surface is increased, creating concentration gradients in a boundary layer between the bulk solution and the membrane surface. This phenomenon of concentration polarization (CP) is a crucial parameter in RO systems because it leads to a larger salt concentration difference across the membrane and an increase in the osmotic pressure of the solution on the surface of the membrane. In addition, it promotes scaling and ``cake layer'' development on the membrane surface. These  combined effects eventually result in a reduction of performance of the RO system. 

% To describe CP , it is customary to use the CP modulus, which is simply the ratio of $c_{i,\beta}/c_{i,feed}$ \cite{Baker2012}.

\section{Model equations}

To model ion rejection and water recovery by RO membranes, appropriate transport models must be set up. Traditionally, there have been two important mechanistic models to describe membrane transport processes: The solution-diffusion (SD) model and the fine capillary pore (FCP) model~\cite{Rautenbach1989}. In the SD model the three transport steps of adsorption, diffusion and desorption are each considered. The case of separating water from dissolved salts results in simplified transport equations due to additional assumptions such as constant water (solvent) concentration in the membrane and equal total pressure on the boundary between membrane and feed side. The SD model provides two membrane parameters which can be determined experimentally, and thus the SD approach offers a method to estimate and plan an RO process. Yet this model simplifies the actual transport mechanisms extensively and does not describe the complexity of the system and acting forces in detail. Instead, the FCP model approaches the film layer as a porous layer with pores of a certain size through which the hydrated ions move. The support layer can also be described in this way, only with much larger pores, but is usually neglected in the theory.  %The solutes move within the pores with a hydration shell. The pores as well are considered to be coated with a layer of water molecules. Since the size of the hydrated ions in the pores, and the pores themselves are comparable, the thicker the hydration layer the better the rejection. 
%The more detailed approach of the FCP model enables calculation of the membrane constants [check]. Nowadays, both for the SD model and the FCP model, there are many different adaptations available.

\begin{figure}[H]
\centering
\includegraphics[width=0.9\textwidth]{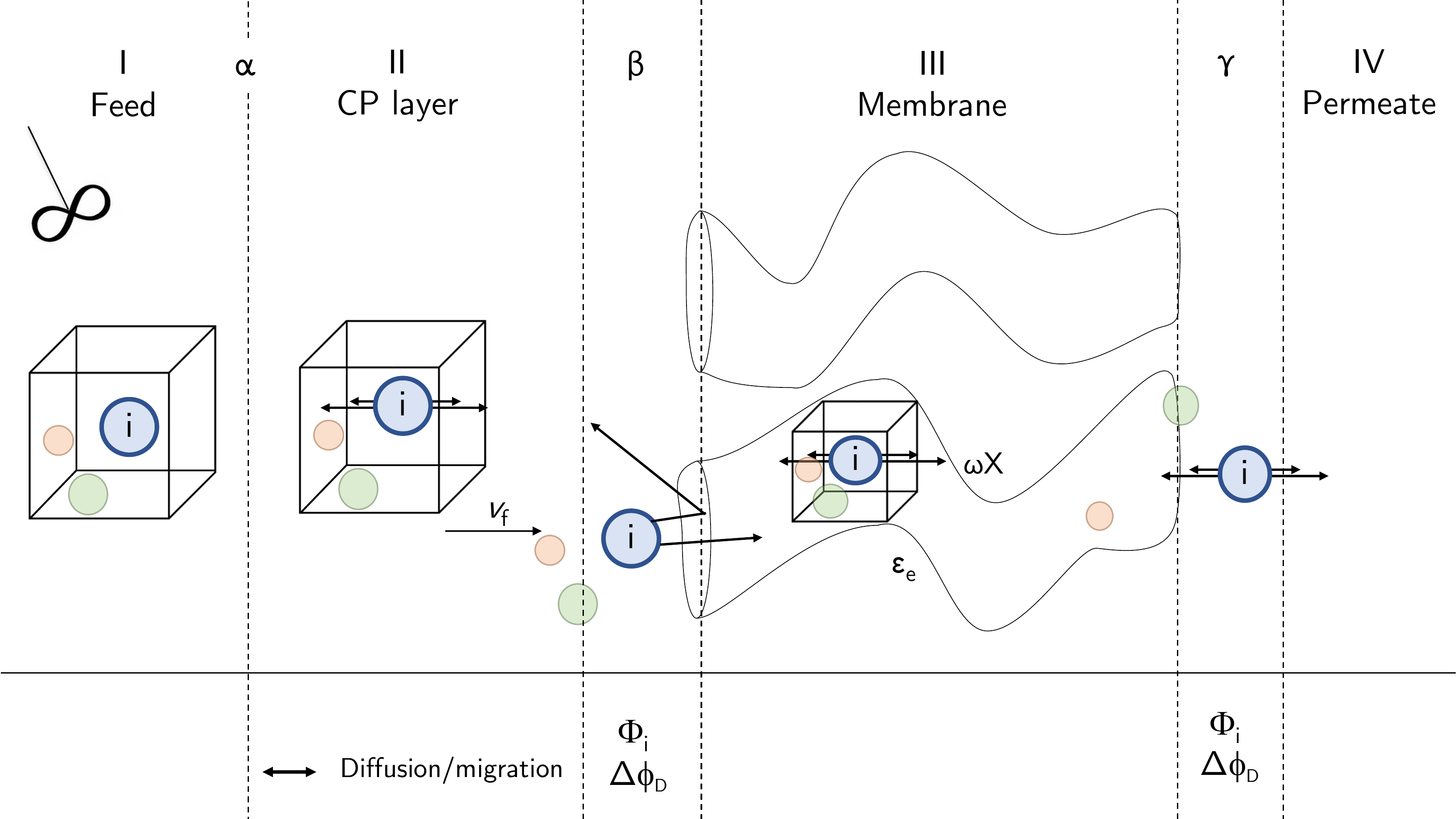}
\caption{Description of the reverse osmosis model used in this work. $\alpha$, $\beta$ and $\gamma$ represent the three boundaries in the system: feed-CP layer, CP layer-membrane and membrane-permeate.} 
\label{fig:Model}
\end{figure}

A general description of our modeling framework is presented in Figure \ref{fig:Model}. Like in the FCP model, in the present work we describe the active layer (III) as a porous material with tortuous pores of a uniform pore size which is of the same order (but larger than) the size of the solutes, thus hindering ion movement. Transport is being described with a Maxwell-Stefan approach, considering three contributions to ion transport: concentration gradients (diffusion), potential gradients (electromigration), and advection of ions with the flow of water.  
For uncharged species, such as boric acid and carbonic acid, there is no migration term involved and their transport is governed by diffusion and advection only. We will neglect friction between solutes, but for all ions include their friction with the fluid (the ``free'' water) as well as with the membrane. Steric partitioning and Donnan effect are included at both membrane-feed interfaces ($\beta$,$\gamma$). The phenomenon of concentration polarization in a layer in front of the membrane is included on the membrane-feed side (II).

%\section{Molecular transport}
%We aim to describe the simultaneous flow of the components in the water matrix through porous (charged) media, such as the selective layer of an RO-membrane. 

%For the moieties we use Maxwell-Stefan theory, which describes the velocity of species i as a balance between driving forces acting on this type of ion, and friction between i with other species (j). The other species can be the fluid (the water in between the ions), other types of ions, and membrane pore walls (the porous matrix).

\subsection{Ion and molecular flux}

We aim to describe the simultaneous flow of the components in the water matrix through a realistic membrane with a certain porosity and tortuosity of the pores. 
The derivation start with a description of a single pore, one which not necessarily follows the shortest distance across the membrane (the ``direct direction''), but instead follows a path that is longer by a factor of $\tau$. Inside such a pore, Maxwell-Stefan theory results in a relationship between the driving forces acting on an ion, and the frictions that it encounters, given by 
\begin{equation} \label{MS}
-\nabla{\mu}_i = R\s{g}T\sum_{j}\zeta_{i-j}\left(v'_i-v'_j\right)
\end{equation}
where $\zeta_{i-j}$ is the friction factor between species $i$ and $j$, $R\s{g}$ is the gas constant, $T$ temperature and $v'_i$ and $v'_j$ are the velocities of the ions in the pore, following the path of the pore. We neglect friction between different ionic species, and only consider friction of ions with the membrane (m) and with the ``free'' water in between the hydrated ions (which has velocity $v\s{f}$, where subcript ``f'' is for fluid). In that case, Eq.~\eqref{MS} can be written in the $x'$-direction as %If we assume only x direction - no need of partial, no? Can just be d__/dx.
\begin{equation} \label{MS2}
-\frac{1}{R\s{g}T}\frac{\partial{\mu}_i}{\partial x'} = \zeta_{i-\text{m}}\left(v'_i-v'_\text{m}\right)+\zeta_{i-\text{f}}\left(v'_i-v'_\text{f}\right)
\end{equation}
where, since the membrane is static, we have $v'_\text{m} = 0$. We will assume ideal thermodynamic behavior for the molecules, (i.e., neglect volumetric interactions, discussed in ref.~\cite{Spruijt}) and thus the electrochemical potential, ${\mu}_i$, is given by
\begin{equation} \label{mu_i}
{\mu}_i={\mu}_{i,0}+R\s{g}T\left(\ln{c_i}+z_i\phi\right)
\end{equation}
where $c_i$ and $z_i$ are the concentration and valency of species $i$, and $\phi$ is the dimensionless electric potential. We can insert Eq. \eqref{mu_i} in Eq. \eqref{MS2} which results after rearrangement in
\begin{equation} \label{v_i_MS}
v'_i = \left(\frac{\zeta_{i-\text{f}}}{\zeta_{i-\text{f}}+\zeta_{i-\text{m}}}\right)v'_\text{f}-\frac{1}{\zeta_{i-\text{f}}+\zeta_{i-\text{m}}}\left(\frac{\partial\ln{c_i}}{\partial x'}+z_i\frac{\partial \phi}{\partial x'}\right).
\end{equation}

Eq. \eqref{v_i_MS} shows structural similarity to hydrodynamic theory for the hindered transport of spherical particles in a pore of comparable size \cite{Deen1987a}. Utilizing this similarity, Eq.~\eqref{v_i_MS} can be written as~\cite{Szymczyk}
\begin{equation} \label{v_i_Deen}
v'_i = K_{\text{c},i} v'_\text{f}-K_{\text{d},i} D_{\infty,i}\left(\frac{\partial\ln{c_i}}{\partial x'}+z_i\frac{\partial \phi}{\partial x'}\right)
\end{equation}
where $K_{\text{c},i}$ and $K_{\text{d},i}$ are called hindrance factors for convection and diffusion. These factors are only a function of the ratio between the diameter of the ion and the pore \cite{Brenner1977,Deen1987a}. Comparing Eqs.~\eqref{v_i_MS} with \eqref{v_i_Deen} shows that
\begin{equation} \label{f_{i-f}}
\zeta_{i-\text{f}} = \frac{K_{\text{c},i}}{K_{\text{d},i} D_{\infty,i}}.
\end{equation}

Note that up to now we have only considered transport within a pore. However, we need to develop a model for fluxes in an actual membrane where part of the structure is not accessible to the ions and water, and in addition, where the paths of the pores are not straight but tortuous (i.e., longer than the shortest direction across the membrane). To account for this, we first implement $dx'=\tau \, dx$, with $\tau$ tortuosity, where $x$ is the coordinate following the shortest distance across the membrane (``direct direction''), and $x'$ is a coordinate along the actual path in the tortuous pore. This modification corrects for the fact that in a pore that is at an angle to the direct direction across the membrane, the average driving force is lower. Furthermore, velocities $v'$ used above are defined along the direction of the tortuous pore, and relate to a superficial velocity $v$ (per unit total membrane area, in the direct direction) by $v=v'\epsilon/\tau$ where $\epsilon$ is membrane porosity. Inserting these relations between $x'$ and $x$, and between $v'$ and $v$, leads to
\begin{equation} \label{eq:v_i}
v_i = K_{\text{c},i}v_f-K_{\text{d},i}\varepsilon\s{e} D_{\infty,i}\left(\frac{\partial\ln{c_i}}{\partial x}+z_i\frac{\partial \phi}{\partial x}\right)
\end{equation}
where $\varepsilon\s{e}=\epsilon/\tau^2$, %
and multiplying Eq. \eqref{eq:v_i} with $c_i$ results for the flux of ion type $i$ in
\begin{equation} \label{J_i}
J_i = K_{\text{c},i}c_i v\s{f}-K_{\text{d},i}\varepsilon\s{e} D_{\infty,i} \left(\frac{\partial c_i}{\partial x}+z_i c_i\frac{\partial \phi}{\partial x}\right).
\end{equation}

\subsection{Water flow}

Also for the the water (fluid) in the pore, we must set up a Maxwell-Stefan-based expression relating fluid velocity $v\s{f}$ to velocities of ions and driving forces acting on the water. %
For the water, the required expression is closely related to Eq.~\eqref{MS}, which in the above section was used to describe the flux of ions. Eq.~\eqref{MS} is a force balance on one ion (one mole of ions) which has friction with the water (filling all space between the ions), pore walls and other ions. For water, we set up a balance per pore volume, $V$, thus the left-hand side of Eq.~\eqref{MS} becomes $-\nabla{\mu}_w \, c\s{w} \, \left(1-\eta\right) \, V$ where $c\s{w}$ is the concentration of water (in water) which is given by $c\s{w}=1/V\s{w}$ where $V\s{w}$ is the molar volume of a water molecule, and where $\eta$ is the volume fraction of all (hydrated) ions in the pore (i.e., the volume excluded for ``free'' water), which we set to zero from this point onward. 

For the friction exerted on the water that is in a volume $V$, we start with the expression on the right-hand side of Eq.~\eqref{MS}, multiplied again by $V$ and for the water-ion frictions multiplied by the concentration of ions, {not of the water}. This can be understood from the fact that the right-hand side of Eq.~\eqref{MS} describes the friction of one ion (or one mole of ions) with the continuum phase of water (and pore walls, and other ions), but we are now interested in the friction of all water that is in a volume $V$ with all ions present in that volume.  Next, we implement in the resulting equation that for water the chemical potential is given by
\begin{equation} \label{mu_w}
\nabla \mu\s{w} = V\s{w} \, R\s{g} T \, \nabla P^t
\end{equation}
where the total pressure $P^t$ is given by \cite{Peters2016}
\begin{equation} \label{P_tot}
P^t = P^h - \Pi
\end{equation}
where $P^h$ and $\Pi$ are the hydraulic and osmotic pressure, both in units of mol/m$^3$ (meaning, pressure in Pa, divided by $R_\mathrm{g}T$). For ideal molecules, i.e., without volume effects, the osmotic pressure is a summation of species concentrations (excluding water molecules), multiplied by $R\s{g}T$, $\Pi=R\s{g}T\,\sum_i c_i$~ \cite{Tedesco2016,Peters2016,Biesheuvel2011}. 
Finally, we include a term for friction between water and pore walls, and then arrive at \cite{Tedesco2016} 
\begin{equation} \label{P_tot_v_f}
\frac{\partial P^t}{\partial x'} = -f'_\text{f-m} v'_\text{f}+\sum_{i} \zeta_{i-\text{f}} c_i \left( v'_i-v'_\text{f} \right)
\end{equation}
where the first term relates to the friction of water (fluid) with the pore walls, and the summation on the right-hand side runs over all ions (not water). The parameter $f'_\text{f-m}$ is the fluid-membrane friction coefficient (defined for the pore) in units of mol$\cdot$s/m$^5$.

In the above derivations for transport of ions and free water in the pore, we have assumed that the ions occupy no volume. This assumption is (implicitly) made at various points, e.g., when we equate velocity $v'_\text{f}$ in Eq.~\eqref{MS2} with that in Eq.~\eqref{v_i_Deen}, and when we neglect the effect of ion volume in Eq.~\eqref{mu_i} and in the derivation of Eq.~\eqref{P_tot_v_f}. One question is at which points in the derivation we must use for the fluid velocity the interstitial fluid velocity, i.e., the actual velocity of the free water in the space left open by the ions, and where a superficial (but still pore based) velocity is required. In future work, the volume occupied by ions in the pore should be included in more detail in the transport equations, describing how it affects the force balance for water, and the recalculation of $v\s{f}$, but in the present work we assume in the transport part of the model that the ion volume fraction is set to $\eta=0$ (ion volume does play a role in the calculation of partition coefficients and hydrodynamic hindrance functions).

Inserting Eqs.~\eqref{v_i_Deen} and \eqref{f_{i-f}} in Eq.~\eqref{P_tot_v_f}, and making the conversions from $x'$ to $x$ and $v'$ to $v$, results in
\begin{equation} \label{P_h}
- \frac{\partial P^t}{\partial x} =  \frac{1}{\varepsilon\s{e}}\left(f'_\text{f-m}+\sum_{i}c_i \zeta_{i-\text{f}} \left(1-K_{\text{c},i}\right)\right) v\s{f} + K_{\text{c},i} \left(\frac{\partial }{\partial x} \sum_i c_i - \omega X \frac{\partial \phi}{\partial x}\right)
\end{equation}
where we implemented the local electroneutrality condition in the membrane
\begin{equation} \label{EN}
\sum_i{z_ic_i}+\omega X = 0 
\end{equation}
where $\omega X$ is the membrane charge density, to be discussed in detail in the next section. By setting $K_{\text{c},i}=1$ (i.e., no ion-wall friction), Eq.~\eqref{P_h} results in the classical expression~\cite{Sonin}
\begin{equation} \label{simple_P_h}
\frac{\partial P^h}{\partial x}+{f\s{f-m}} v\s{f} = \omega X \frac{\partial \phi}{\partial x}
\end{equation}
where $f\s{f-m}=f'_\text{f-m}/\varepsilon\s{e}$. 

%
%which is the same expression obtained by %Sonin [40] and Schlogl [42] from Tedesco2016. 

In the following sections we describe further elements of the model, relating to the PA toplayer in the membrane, pore and ion sizes, partitioning at membrane edges, hindrance factors and membrane charge.

\subsection{Effective ion and pore sizes in PA membrane}

For the TFC SWRO membrane that we model, it was assumed that the main resistance to transport in the membrane derives from the polyamide layer and therefore the membrane description is of that layer alone. The layer was considered to be a rigid structure \cite{Shen2016,Freger2004,Kosutic2006} and thus we chose a defined pore size. An average pore size of 0.76 nm~\cite{Kezia2014,Kosutic2006} and film thickness of 100 nm were assumed. These values are within the range of values reported in literature~\cite{Coronell2008,Ghosh2008,Kezia2014,Kosutic2006,Shen2016,
Yoon2005,Zhang2007}. Owing to the aqueous environment in the membrane, species were described with their solvation shell included. However, since the hydrated size of carbonate and bicarbonate exceeds the value for pore size assumed, it was considered that they have to, at least partially, shed their solvation shell, and thus for the carbonate species radii were chosen to represent partially hydrated ions yielding $\lambda_i < 1$. All ion sizes used in the model are summarized in Table \ref{Inputs}.

\subsection{Donnan-steric partitioning}

As a result of membrane charge, an electrochemical potential develops on both membrane edges (at the bulk--membrane interfaces, $\beta$ and $\gamma$, see Fig.~\ref{fig:Model}). This phenomenon is known as the Gibbs-Donnan effect and the electric potential difference as Donnan potential. Based on the calculated value of the Donnan potential we can relate the concentration of the ions between both sides of the boundary.
It has been suggested that for the ions to be able to enter the pores they have to, at least partially, loose or rearrange their solvation shell which in turn poses an energy barrier for the process \cite{Coronell2008,Tansel2012}. However, in the present work,  we only account for the steric disturbance, by introducing a partitioning coefficient at the membrane-solution interface, described by
\begin{equation} \label{partition}
\Phi_i = \left(1-\lambda_i\right)^2
\end{equation}
where $\lambda$ is the ratio between the ion size and the pore size. %Note that this correction should be applied only in cases where $\lambda_i<1$. 
Combining the partitioning and electrostatic effects, we obtain 
\begin{equation} \label{Donnan-partition}
c_{i,\text{m}} = c_{i,\infty}\Phi_i\exp(-z_i\Delta\phi\s{D})
\end{equation}
where $\Delta\phi\s{D}$ is the dimensionless Donnan potential (Donnan potential divided by $R_\mathrm{g}T/F$) and subscript m refers to a point just within the membrane, and $\infty$ to just on the outside. 

\subsection{Membrane charge} \label{subsection:Mem_charge}

Coronell \textit{et al}. quantified the charged functional groups in an FT30 RO membrane \cite{Coronell2008}. They found that while ionized amine groups can be described by one dissociation constant, pK$_{\ce{NH2}}$, two values were required to adequately describe the carboxylic groups, pK$_{\ce{COOH1}}$, pK$_{\ce{COOH2}}$. The equilibrium for the carboxylic groups is given by
\begin{equation} \label{EQ1}
{[\ce{R-COOH}]}\rightleftharpoons[\ce{H3O+}]+[\ce{RCOO-}], \ \ \ K_{\ce{COOH}}
\end{equation}
where $R-$ represents the polymer backbone. Note that throughout this work, all $K$ and pK-values for chemical equilibria refer to acid constants. Utilizing the relation between concentration and equilibrium constant, $K_{\ce{COOH}}$, we can write
\begin{equation} \label{[R-COOH]_{eq}}
[\ce{RCOOH}] = \frac{[\ce{H3O+}][\ce{RCOO-}]}{K_{\ce{COOH}}}
\end{equation}
and due to the fixed number of carboxylic groups, at all times 
\begin{equation} \label{Cons}
[\ce{RCOOH}]\s{tot}=[\ce{RCOO-}]+[\ce{RCOOH}]
\end{equation}
%
%With $[R-COOH]_{tot}$ being the total possible amount of $COOH$ (''$X_{COOH}$'') and by inserting Eq. \eqref{[R-COOH]_{eq}} into Eq. \eqref{Cons} we now get 
%
and with $X_{\ce{COOH}}=[\ce{RCOOH}]\s{tot}$, we arrive at
\begin{equation} \label{RCOO}
[\ce{RCOO-}] = {X_{\ce{COOH}}}/\left({1+\frac{[\ce{H3O+}]}{K_{\ce{COOH}}}}\right).
\end{equation}

\begin{figure}[H]
\centering
\includegraphics[width=0.5\textwidth]{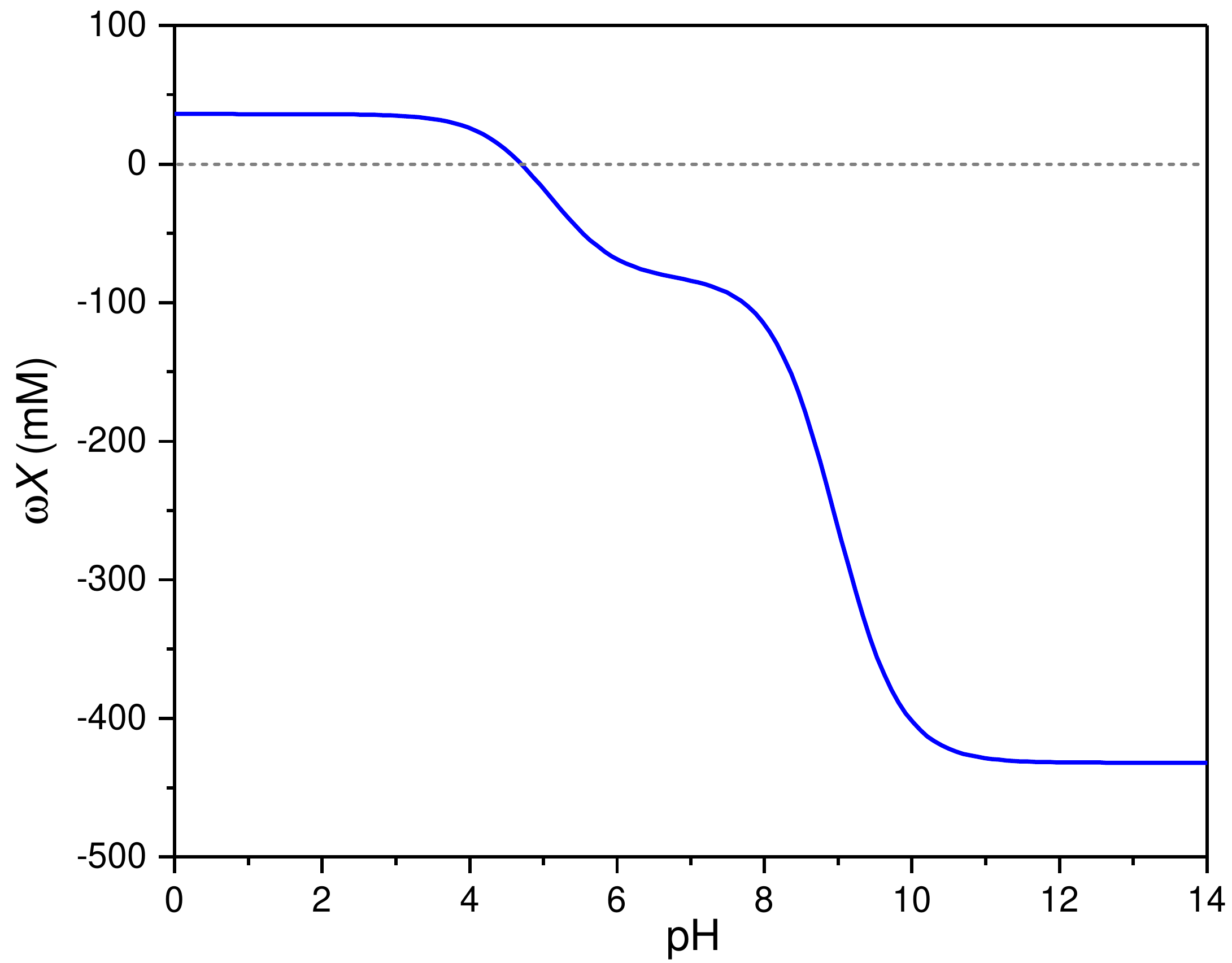}
\caption{Fixed membrane charge in polyamide RO membrane toplayer, as a function of local pH. Calculated from ref.~\cite{Coronell2008}.} \label{MemCharge}
\end{figure}

Similarly, for the amine surface groups, equilibrium is given by
\begin{equation} \label{EQ2}
{[\ce{RNH3+}]}\rightleftharpoons[\ce{H3O+}]+[\ce{RNH2}]
\end{equation}
and considering conservation of amine groups, the expression for the protonated amine group is
\begin{equation} \label{RNH2}
[\ce{RNH3+}] = {X_{\ce{NH2}}}/\left({1+\frac{K_{\ce{NH2}}}{[\ce{H3O+}]}}\right).
\end{equation}
Assuming fast equilibrium and taking into account both carboxylic groups, we use Eqs. \eqref{RCOO} and  \eqref{RNH2} to write
\begin{equation} \label{X-pH}
\omega X = {X_{\ce{NH2}}}/\left({1+\frac{K_{\ce{NH2}}}{[\ce{H3O+}]}}\right)-{X_{\ce{COOH1}}}/\left({1+\frac{[\ce{H3O+}]}{K_{\ce{COOH1}}}}\right)-{X_{\ce{COOH2}}}/\left({1+\frac{[\ce{H3O+}]}{K_{\ce{COOH2}}}}\right)
\end{equation}
which describes the membrane charge at any given \ce{H3O+} concentration in the PA layer, i.e. for any local pH. The result of Eq.~\eqref{X-pH} is plotted as function of pH in Fig. \ref{MemCharge}, based on data/input of ref.~\cite{Coronell2008}. Though the membrane charge as measured in ref.~\cite{ Coronell2008} is determined as a concentration per unit total volume of the toplayer, in the present work we use their result as if it is the membrane charge per unit pore volume. 

\subsection{Hindered transport}

Since the pore molecular size is not much larger than that of the ions, it is important to include the pore size in the transport model. Here we rely on hindered transport theory, as developed and discussed in literature~\cite{Brenner1977,Deen1987a}. The expressions for the steric hindrance coefficients divide into two ranges. For $0<\lambda_i<0.8$ the expression was adapted from Bowen \textit{et al}. \cite{Bowen1997} and for $0.8<\lambda_i<1$ from Bandini and Vezzani~ \cite{Bandini2003}, which are given by
\begin{equation}
\begin{aligned}
K_{\text{c},i} &= \left\{
  \begin{array}{lr}
    1.0+0.054\lambda_i-0.988\lambda_i^2+0.441\lambda_i^3 & 0 < \lambda_i < 0.8\\
    -6.83+19.348\lambda_i-12.518\lambda_i^2 & 0.8 <\lambda_i <1
  \end{array}
\right.\\
K_{\text{d},i} &= \left\{
  \begin{array}{lr}
    1.0-2.30\lambda_i+1.154\lambda_i^2+0.224\lambda_i^3 & \;\;\;0 < \lambda_i < 0.8\\
    -0.105+0.318\lambda_i-0.213\lambda_i^2 &  0.8 <\lambda_i < 1
  \end{array}
\right.
\end{aligned}
\end{equation}

The values for $\lambda_i$, $K_{\text{c},i}$ and $K_{\text{d},i}$ used in the model are presented in Table~\ref{Inputs}.

\subsection{Concentration polarization (CP) layer}

For the description of the CP layer in front of the membrane, the same transport equations were used as presented above, but now excluding hindered transport, as well as porosity and tortuosity effects, and thus ionic flux in the CP layer is described by the Nernst-Planck equation,
\begin{equation} \label{J_i_cp}
J_i = c_i v\s{f}-D_{\infty,i}\left(\frac{\partial c_i}{\partial x}+z_i c_i\frac{\partial \phi}{\partial x}\right)
\end{equation}
while Eq.~\eqref{simple_P_h} results in  the fact that in the CP-layer the hydrostatic pressure gradient is zero. The model considers a CP layer thickness of $\delta\s{CP}=20~\mu$m \cite{Baker2004}. Electroneutrality in the CP layer is given by Eq.~\eqref{EN} with $\omega X=0$.

\subsection{Auxiliary relations for ionic fluxes}

Several additional relations are required to obtain a complete model. First of all, in RO the ionic current, $J\s{ch}$, is zero, and thus the summation of all ion fluxes (times valency) must be zero at each point in the CP layer, as well as at each point in the membrane. Thus, at each point
\begin{equation} \label{current_zero}
J\s{ch} = \sum_i{z_i J_i} = 0.
\end{equation}

Until now we have not yet explicitly described how the acid/base reactions play a role in the structure of the model. As mentioned in section \ref{seawater_characteristics}, transport of weak acid systems, such as boric and carbonic acid, is more complex to describe because of the local chemical equilibrium that is affecting concentrations. This is in contrast to strong acid systems where the equilibrium constant is large enough to neglect changes that occur due to a disturbance of chemical equilibrium. In our model, we take into consideration that all species, except for sodium and chloride, relate to each other through a reaction involving the hydronium ion. This means that a change in the concentration of for instance \ce{HCO3-}, will actuate a change in concentration of \ce{H2CO3} and \ce{HCO3-} and of \ce{H3O+}, \ce{OH-}, but also of \ce{B(OH)3} and \ce{B(OH)4-}. What this implies is that in the mathematical code, because we work at steady-state, the flux not of an individual ion, like \ce{HCO3-}, is invariant across the membrane (and the same at any point in the CP layer), but only the flux of the \textit{group} of the three carbonate species together~\cite{Dykstra2014}. And the same holds for the group of the two boron species. Note that it is not necessary to set up a flux equation for \ce{H3O+} and \ce{OH-}~\cite{Dykstra2014}. The diffusion coefficient of these species only shows up in the expression for the ionic current, Eq.~\eqref{current_zero}.

Finally, we need to relate the concentration in the permeate to the fluxes through the membrane. For the dead-end cell experiments we consider, the following relationship then holds, where
\begin{equation} \label{dead end}
c_{i,\text{permeate}}=J_i / v\s{f}.
\end{equation}

It is important to note that in this form, Eq.~\eqref{dead end} can only be used for \ce{Na+} and \ce{Cl-}. For the carbonate and borate groups, Eq.~\eqref{dead end} is used for the flux and concentration of the entire group. Eq.~\eqref{dead end} should not be used for \ce{H3O+} and \ce{OH-}. 

All equations are discretized and solved in steady-state, for a given feed composition, and given fluid flow rate, $v\s{f}$. The hydrostatic pressure difference, $\Delta P^h$, required to achieve a given flow rate, is calculated afterwards.

\section{Results and Discussion}

\subsection{Input parameters in the model}

In this section we will present results of calculations for one given feed seawater composition, and one set of input parameters for ion size, diffusion coefficients, etc. The only parameters that will be changed are the water flow rate through the membrane, $v\s{f}$, membrane charge, and seawater feed pH. All input parameters used in the model are listed in Table \ref{Inputs}. Note that in this section (and throughout this paper), the words ``membrane'' and ``PA toplayer'' are both used, and have the same meaning. The same goes for the words (``free'') water and ``fluid''.

\begin{table}
\centering
\caption{Inputs used in the simulation.\\$\ast$ Based on Stokes-Einstein relation using $\mu\s{w}=8.9\cdot 10^{-4}$ Pa.s \cite{Goli2010}.\\$\dagger$ Partial loss of hydration shell was assumed.\\$\ddagger$ based on $f\s{f-m} = \left(A\delta\s{m} R\s{g}T\right)^{-1}$ \cite{Tedesco2016}, with $A=3.0~\mu$m/bar/s \cite{Ghosh2008}.}

\label{Inputs}
\begin{adjustbox}{center}
\begin{small}
\begin{tabular}{lllllll}
\toprule
    & Parameter     & Value       &  & Parameter       & Value         \\
\midrule
 \multirow{7}{*}{ \begin{turn}{+90}constituents\end{turn} }    	&               				&             		&    \multirow{7}{*}{ \begin{turn}{+90}equilibrium\end{turn}} 	& pK\textsubscript{\ce{COOH1}}       & 5.23 \cite{Coronell2008}     \\
										& \ce{Na+}\textsubscript{feed}	& 553\,mM 		& 										& pK\textsubscript{\ce{COOH2}}        & 8.97 \cite{Coronell2008}     \\
              									&  \ce{Cl-}\textsubscript{feed}  	&  550\,mM       	&  										& pK\textsubscript{\ce{NH3}}         & 4.74 \cite{Coronell2008}  \\
               									&  Btot\textsubscript{feed}     	&    0.5\,mM \cite{Company,Koseoglu2008}           	&  										& pK\textsubscript{\ce{B(OH)3}}       & 8.60 \cite{Dickson1987} \\ 
               									&     Ctot\textsubscript{feed}  	&  2.48\,mM \cite{Fritzmann2007,Lenntecha}         	&  										& pK\textsubscript{\ce{H2CO3}}       & 5.98 \cite{Dyrssen1973}  \\
               									&       pH\textsubscript{feed}     	&   8.0 \cite{Fritzmann2007,Company}          	&  										& pK\textsubscript{\ce{HCO3-}}      & 9.16 \cite{Dyrssen1973}      \\
                              								&               				&             		&  										& pK\textsubscript{w}                       & 13.3 \cite{Dyrssen1973}     \\
               \hline
\multirow{9}{*}{ \begin{turn}{+90}properties\end{turn} }    & $d$\textsubscript{\ce{Na+}} & 7.16 \AA \cite{Nightingale1959}  &  & $d$\textsubscript{\ce{Cl-}}   & 6.64 \AA \cite{Nightingale1959}        \\
               & $d$\textsubscript{\ce{H3O+}}  & 5.64 \AA \cite{Nightingale1959}         &  & $d$\textsubscript{\ce{OH-}}   & 6.00 \AA        \cite{Nightingale1959} \\
               & $d$\textsubscript{\ce{B(OH)3}} & 3.84 \AA\textsuperscript{*}            &  & $d$\textsubscript{\ce{B(OH)4-}} & 5.22 \AA \cite{Corti1980}     \\
               & $d$\textsubscript{\ce{H2CO3}} & 3.64 \AA \textsuperscript{*}            &  & $d$\textsubscript{\ce{HCO3-}} & 7.16 \AA \textsuperscript{$\dagger$}      \\
               & $d$\textsubscript{\ce{CO3^2-}} & 7.30 \AA \textsuperscript{$\dagger$}            &  & $d$\textsubscript{pore}  & 7.60 \AA \cite{Kezia2014,Kosutic2006}\\      
               & $D$\textsubscript{\ce{H3O+}}& \SI{8.24e-09}{\metre^{2}\per\s} \cite{Lee2011}   &  & $D$\textsubscript{\ce{OH-}} & \SI{4.51e-09}{\metre^{2}\per\s} \cite{Lee2011}    \\
               & $D$\textsubscript{\ce{Na+}}          & \SI{1.33e-09}{\metre^{2}\per\s} \cite{Atkins2009}   &  & $D$\textsubscript{\ce{Cl-}} & \SI{2.00e-09}{\metre^{2}\per\s} \cite{Atkins2009}\\
               & $D$\textsubscript{\ce{B(OH)3}}          & \SI{1.28e-09}{\metre^{2}\per\s} \cite{Goli2010}   &  & $D$\textsubscript{\ce{B(OH)4-}}          & \SI{1.18e-09}{\metre^{2}\per\s}    \\
               & $D$\textsubscript{\ce{H2CO3}}         & \SI{1.92e-09}{\metre^{2}\per\s} \cite{Dykstra2014}   &  & $D$\textsubscript{\ce{HCO3-}}          & \SI{1.18e-09}{\metre^{2}\per\s} \cite{Dykstra2014}  \\
               & $D$\textsubscript{\ce{CO3^{2-}}}     & \SI{9.8e-10}{\metre^{2}\per\s} \cite{Dykstra2014}    &  &                 &             \\          
               \hline
\multirow{4}{*}{ \begin{turn}{+90}membrane\end{turn} }  & $\delta\s{m}$           & 100 nm \cite{Kezia2014,Coronell2008,Ghosh2008}  &  & $\delta\s{CP}$            & 20 $\mu$m \cite{Baker2004}    \\
               & $\varepsilon\s{e}$           & 0.05 \cite{Kezia2014}       &  &$f$\textsubscript{f-m}            & \SI[per-mode=symbol]{1.3d14}{\mole.\s\per\metre^{5}} \textsuperscript{$\ddagger$}   \\
               & $X$\textsubscript{\ce{COOH1}}          & 82\,mM \cite{Coronell2008}         &  & $X$\textsubscript{\ce{COOH2}}            & 350\,mM \cite{Coronell2008}        \\
               & $X$\textsubscript{\ce{NH3}}             & 36\,mM \cite{Coronell2008}         &  &          &       \\
               %&               &             &  &                 &           \\
               \hline
\multirow{10}{*}{ \begin{turn}{+90}partitioning (calculated)\end{turn} }   & $\lambda$\textsubscript{\ce{Na+}}         & 0.942    &  & $\lambda$\textsubscript{\ce{Cl-}}           & 0.874     \\
               & $\lambda$\textsubscript{\ce{H3O+}}          & 0.742    &  & $\lambda$\textsubscript{\ce{OH-}}           & 0.790     \\
               & $\lambda$\textsubscript{\ce{B(OH)3}}         & 0.505    &  & $\lambda$\textsubscript{\ce{B(OH)4-}}         & 0.687    \\
               & $\lambda$\textsubscript{\ce{H2CO3}}         & 0.479    &  & $\lambda$\textsubscript{\ce{HCO3-}}         & 0.942    \\
                & $\lambda$\textsubscript{\ce{CO3^2-}}         & 0.961    &  &                 &           \\
               & $\Phi$\textsubscript{\ce{Na+}}        & \num{0.00335}    &  & $\Phi$\textsubscript{\ce{Cl-}}          & \num{0.0156}     \\
               & $\Phi$\textsubscript{\ce{H3O+}}         & \num{0.0665}     &  & $\Phi$\textsubscript{\ce{OH-}}          & \num{0.0443}     \\
               & $\Phi$\textsubscript{\ce{B(OH)3}}        & \num{0.245}    &  & $\Phi$\textsubscript{\ce{B(OH)4-}}        & \num{0.0981}    \\
               & $\Phi$\textsubscript{\ce{H2CO3}}        & \num{0.272}    &  & $\Phi$\textsubscript{\ce{HCO3-}}        & \num{0.00335}        \\
              & $\Phi$\textsubscript{\ce{CO3^2-}}        & \num{0.00156}    &  &                 &           \\
               \hline
\multirow{9}{*}{ \begin{turn}{+90}hindrance (calculated)\end{turn} }     & $K$\textsubscript{c,\ce{Na+}}      & 0.287 &  & $K$\textsubscript{d,\ce{Na+}}      & \num{0.00554} \\
               & $K$\textsubscript{c,\ce{Cl-}}        & 0.519 & & $K$\textsubscript{d,\ce{Cl-}}        & \num{0.0102} \\
               & $K$\textsubscript{c,\ce{H3O+}}        & 0.676 &  & $K$\textsubscript{d,\ce{H3O+}}        & \num{0.0202}  \\
               & $K$\textsubscript{c,\ce{OH-}}        & 0.644 &  & $K$\textsubscript{d,\ce{OH-}}        & \num{0.0137}  \\
               & $K$\textsubscript{c,\ce{B(OH)3}}      & 0.832  &  & $K$\textsubscript{d,\ce{B(OH)3}}     & 0.161   \\
               & $K$\textsubscript{c,\ce{B(OH)4-}}      & 0.714 &   & $K$\textsubscript{d,\ce{B(OH)4-}}      & \num{0.0373}  \\
               & $K$\textsubscript{c,\ce{H2CO3}}      & 0.848 &  & $K$\textsubscript{d,\ce{H2CO3}}      & 0.188  \\
               & $K$\textsubscript{c,\ce{HCO3-}}      & 0.287 &  & $K$\textsubscript{d,\ce{HCO3-}}      & \num{0.00554} &    \\
               &  $K$\textsubscript{c,\ce{CO3^2-}}      & 0.205 &              & $K$\textsubscript{d,\ce{CO3^2-}}      & \num{0.00393}  \\
               \hline
\multirow{3}{*}{ \begin{turn}{+90}\end{turn} }%\\
        & $R_\mathrm{g}$            & 8.3144 J/K/mol   & &                 &      \\
                & $F$               & 96845 C/mol      &  &                 &         \\
                & $T$               & 25 $^\circ$C      &  &                 &         \\
               \bottomrule
\end{tabular}
\end{small}
\end{adjustbox}
\end{table}
%**In the model - activities are taken to be 1. \\

\subsection{Results as function of permeate flow rate}

Before investigating the influence of water flow rate, we first present results of calculations using a permeate flow rate (water flow rate through the membrane) of $v\s{f}=10~\mu$m/s, which is a typical value in SWRO, and which recalculates to 36 L/m$^2$/hr~\cite{Baker2004}. The performance of the membrane at this condition in terms of rejection of salt and boron is summarized in Table \ref{Table:results} and compared with values from literature. As shown in Table \ref{Table:results}, we obtain a near-perfect match for all three properties. The model gives a slightly optimistic prediction of salt and boron rejection which suggests that in the calculation the pore size must perhaps be increased slightly. The calculated pH\textsubscript{permeate} is within the range of experimental data reported in ref.~\cite{Andrews2008}. An elevation in pH\textsubscript{permeate} comparing with pH\textsubscript{feed} was also observed in ref.~\cite{Nir2014a} with a feed (seawater) pH of 9.0 resulting in permeate pH of 9.1--9.5. According to Eq.~\eqref{P_h}, for this water flow rate, the required applied pressure was calculated to be $\Delta P^h=35.6$ bar. Based on this value and the osmotic pressure of the feed solution, the thermodynamic efficiency of this process can be calculated as $\eta=77$\% based on $\eta=\Delta\Pi/\Delta P^h$ where the osmotic pressure difference is that across the membrane. Note that this relation is only valid for a very low water recovery and a very dilute permeate, see Eq.~(3) in ref.~\cite{Biesheuvel2009}.

\begin{table}[H]
\centering
\caption{Summary of model output compared with values for sea water reverse osmosis reported in literature.}
\label{Table:results}
\begin{tabular}{llll}
\toprule
Parameter       & Our model                                           & Literature                                               & Reference     \\                                                           \midrule
pH\textsubscript{permeate}      & 8.8                                                & 8.6-8.8                                                  & \begin{tabular}[c]{@{}l@{}}\cite{Andrews2008}\\  

\end{tabular}         \\
\textit{R}\textsubscript{NaCl}  & \begin{tabular}[c]{@{}l@{}}99.9\,\% \\  \end{tabular} & \begin{tabular}[c]{@{}l@{}}96.6-99.8\,\%\\  \end{tabular} & \begin{tabular}[c]{@{}l@{}}\cite{Dominguez-Tagle2011}\\   \end{tabular} \\
\textit{R}\textsubscript{B} & \begin{tabular}[c]{@{}l@{}}93.5\,\%\\   \end{tabular} & \begin{tabular}[c]{@{}l@{}}87-93\,\%\\   \end{tabular}     & \begin{tabular}[c]{@{}l@{}}\cite{Dominguez-Tagle2011}\\ \end{tabular}
\\
\bottomrule
\end{tabular}
\end{table}

Calculation results show that the CP layer increases the salt concentration directly on the membrane surface from $c\s{salt,feed} = 550$ mM in the feed, to about $c\s{salt,\beta} = 620$\,mM. This is about a 10\% increase, which is less than a typical value in RO of around 30\%~\cite{Baker2004}. Thus, in our calculation the CP layer increases the osmotic pressure of the salt solution directly near the membrane surface by $\sim$4 bar.  %
As observed in calculations at different values of $v\s{f}$, for all species but \ce{H3O+}, increasing $v\s{f}$ always results in an increase in ion concentrations at the membrane surface. 

\begin{figure}
\centering
\includegraphics[width=0.8\textwidth]{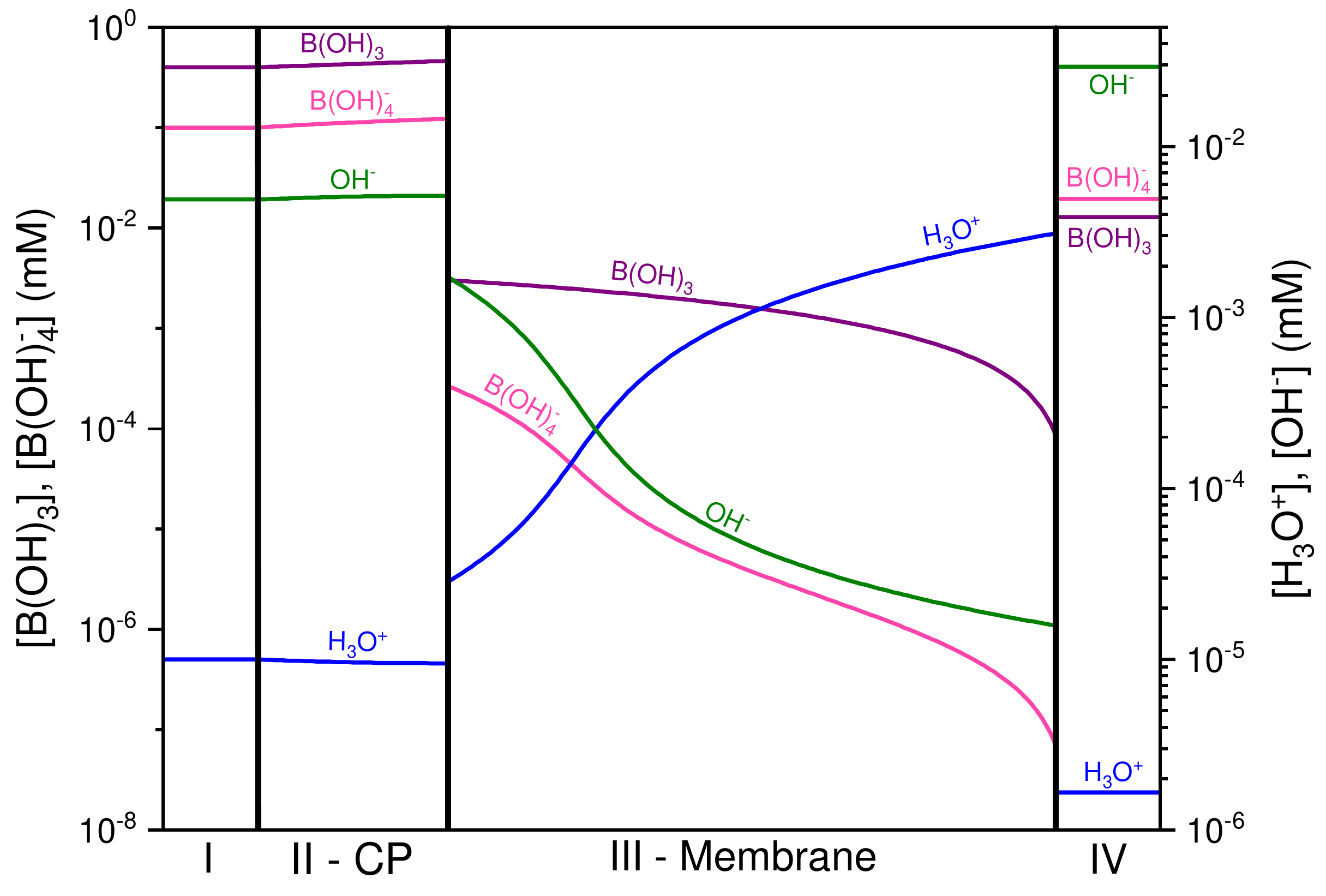}
\caption{Concentration profiles of the two boron species and \ce{H3O+} and \ce{OH-} in the CP-layer (II) and membrane (III). Concentrations are plotted on a logarithmic scale. Left is the feed solution (I), right permeate (IV).} 	  \label{fig:c_system}
\end{figure}

As an example, we show in Fig.~\ref{fig:c_system} concentration profiles in the CP layer and in the membrane for the boric acid and the borate ion, as well as of \ce{H3O+} and \ce{OH-}. For \ce{H3O+} and \ce{OH-}, the product of their concentrations is always the same and therefore their profiles are vertical mirror images. A condition of pH 7 is reached at a point around one third into the membrane. pH decreases by about two points when we move from the very left to the very right of the PA toplayer, and pH then increases by three points when we exit the membrane. In the membrane the neutral boric acid molecule is about an order of magnitude more prevalent than the borate anion.

\begin{figure}[H]
\centering
\includegraphics[width=1\textwidth]{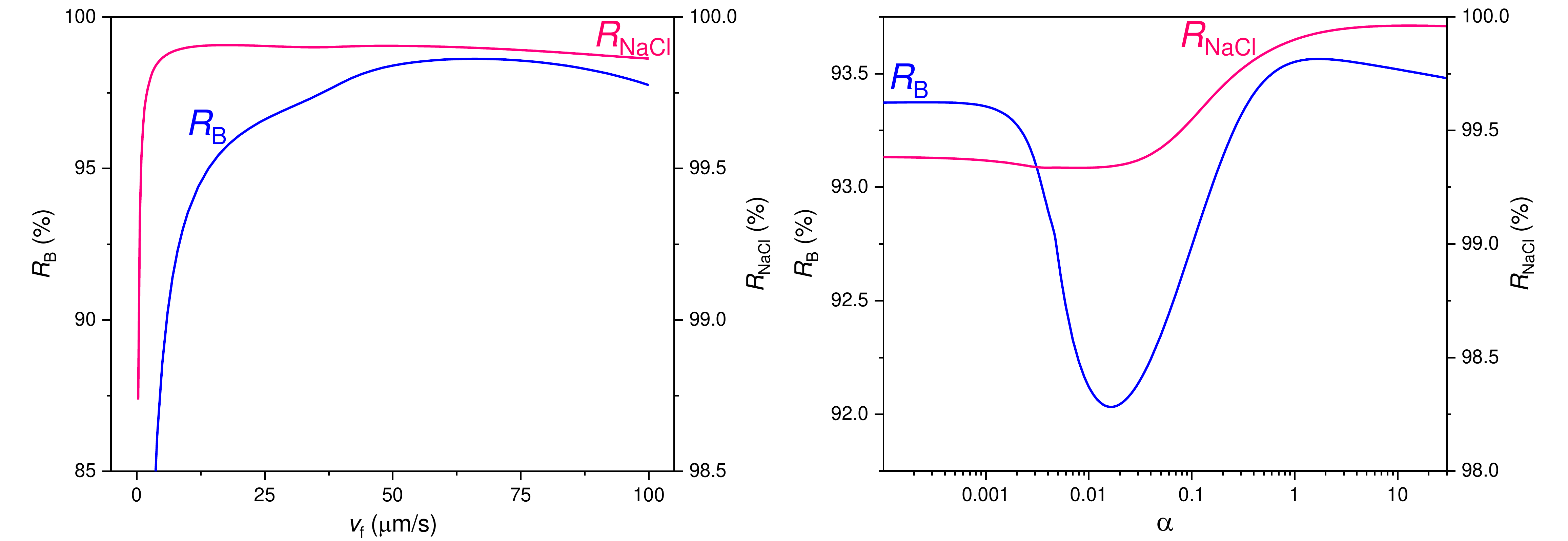}
\caption{Rejection of NaCl and boron by the membrane as function of A) water flow rate $v\s{f}$, and B) membrane charge density $\alpha$, which is the factor by which the concentration of each component of membrane charge, $X\s{k}$, is multiplied. Conditions of Table~\ref{Inputs}, $v\s{f}=10~\mu$m/s.} \label{fig:RBRNa_combined}
\end{figure}

Next we examined the effect of fluid flow rate (through the membrane) on system performance, both for much lower and much higher flow rates than the standard situation just discussed. Salt rejection remained almost constant throughout the examined range at a value between 99.0\% and 99.9\%, see Fig. \ref{fig:RBRNa_combined}A, while boron rejection increased from 93.5\% at $v\s{f}=10~\mu$m/s to $R\s{B}=98.6\%$ at $v\s{f}=60~\mu$m/s. Lower flow rates reduced boron retention significantly. Increasing flowrate from 10 to 40 $\mu$m/s, permeate pH decreased from 8.8 to $\sim8.6$ after which it increases again to reach pH 9.5 at a flowrate of 100 $\mu$m/s [results not shown]. Thus, for all flow rates, pH\textsubscript{permeate} was higher than feed pH.

\begin{figure}[H]
\centering
\begin{subfigure}{.5\textwidth}\label{subfig:c_Na}
\centering
\includegraphics[width=1.0\linewidth]{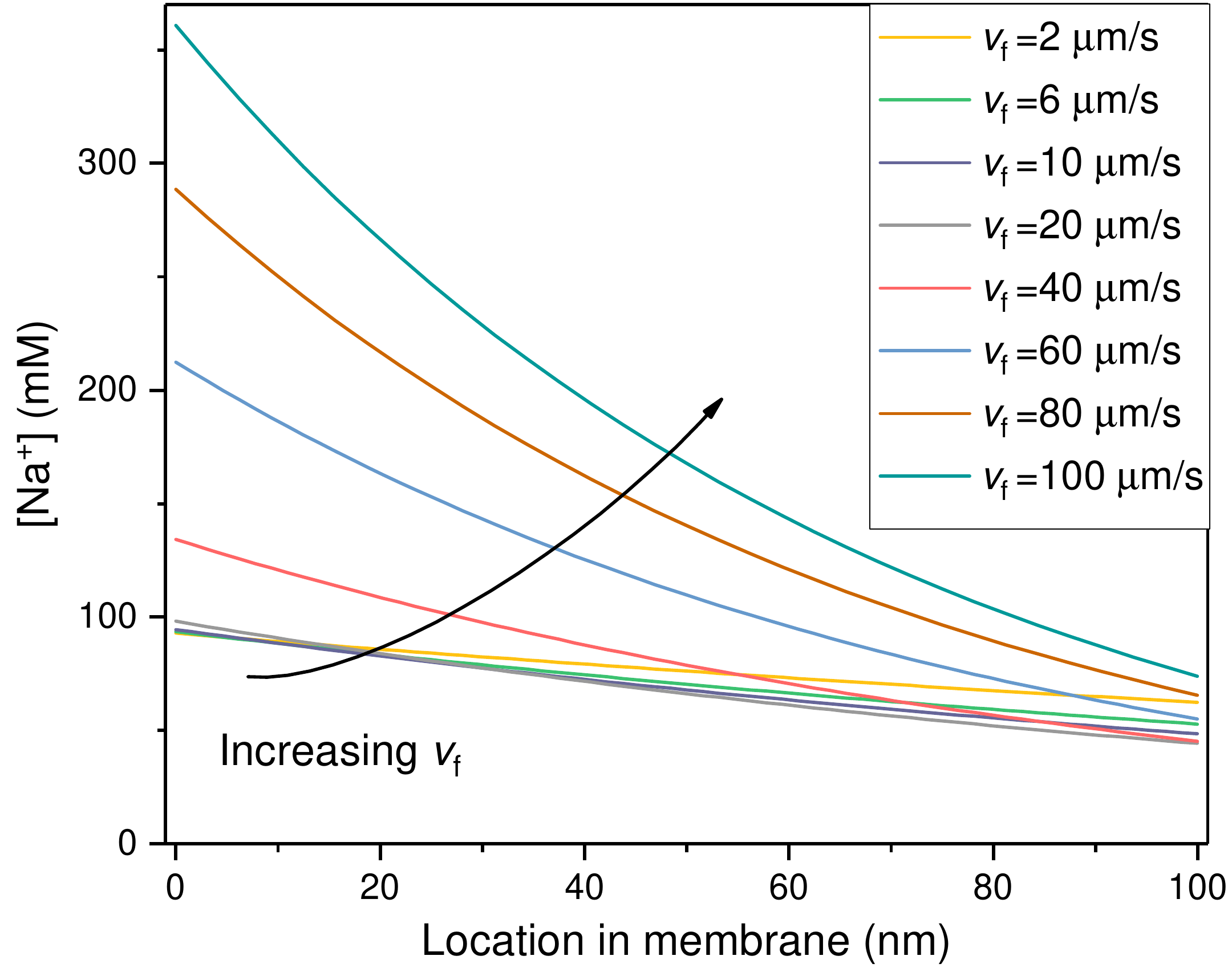}
	  \caption{\ce{Na+}}
	  \end{subfigure}%
  \begin{subfigure}{.5\textwidth}
  \centering
\includegraphics[width=1.0\linewidth]{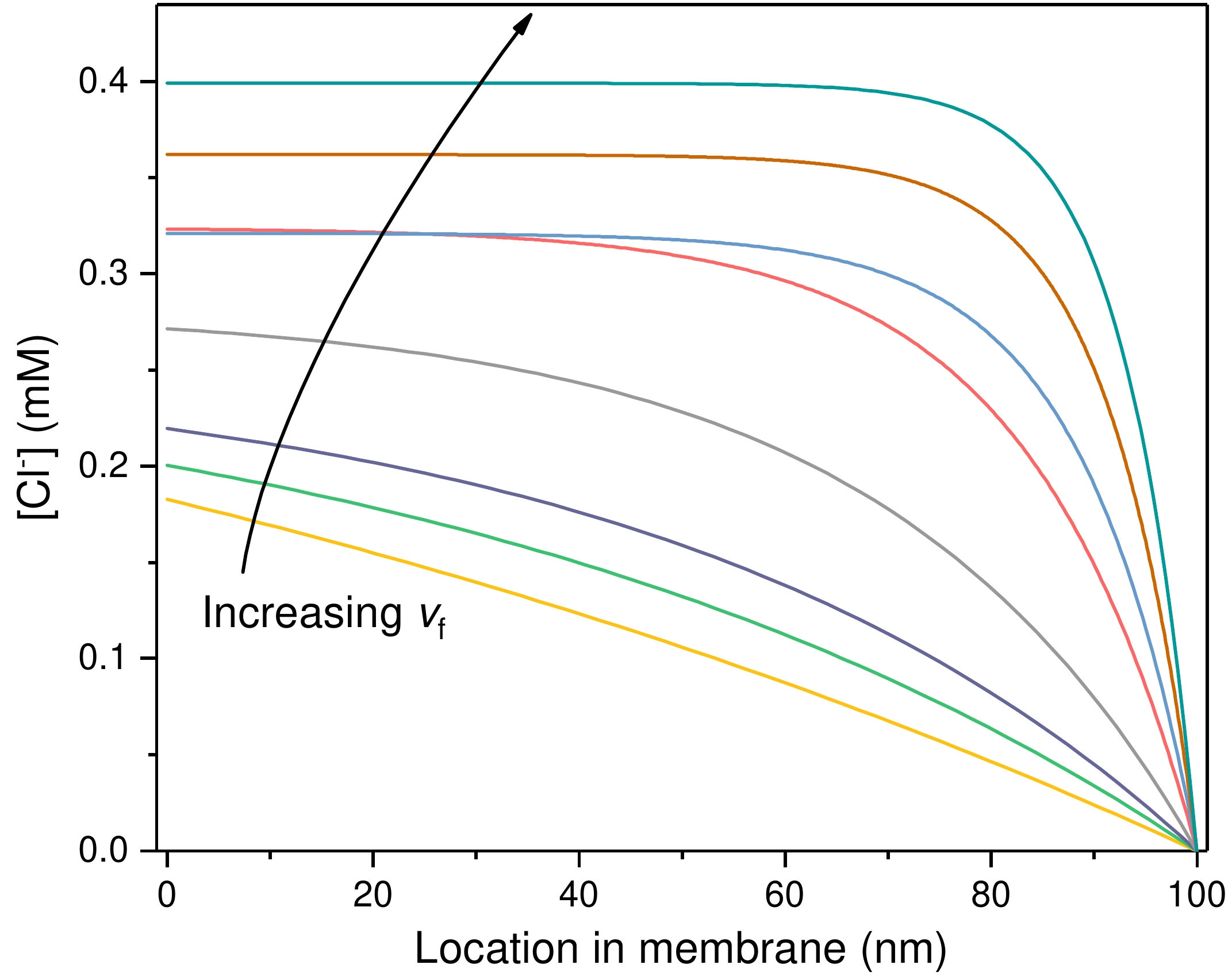}
	  \caption{\ce{Cl-}}
	  \end{subfigure}
	  \begin{subfigure}{.5\textwidth}\label{c_H}
  \centering
\includegraphics[width=1.0\linewidth]{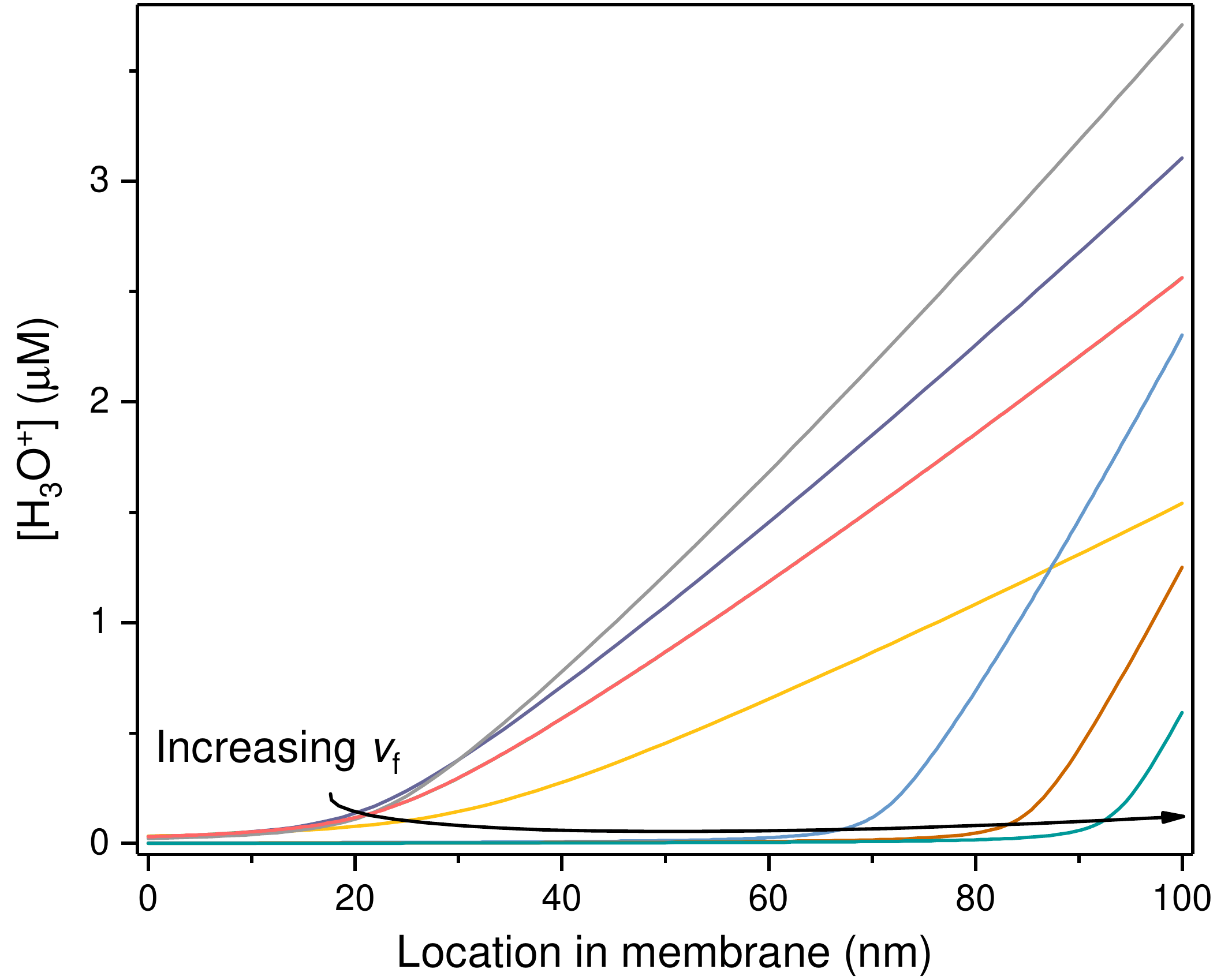}
	  \caption{\ce{H3O+}}
	  \end{subfigure}%
	  \begin{subfigure}{.5\textwidth}
  \centering
\includegraphics[width=1.0\linewidth]{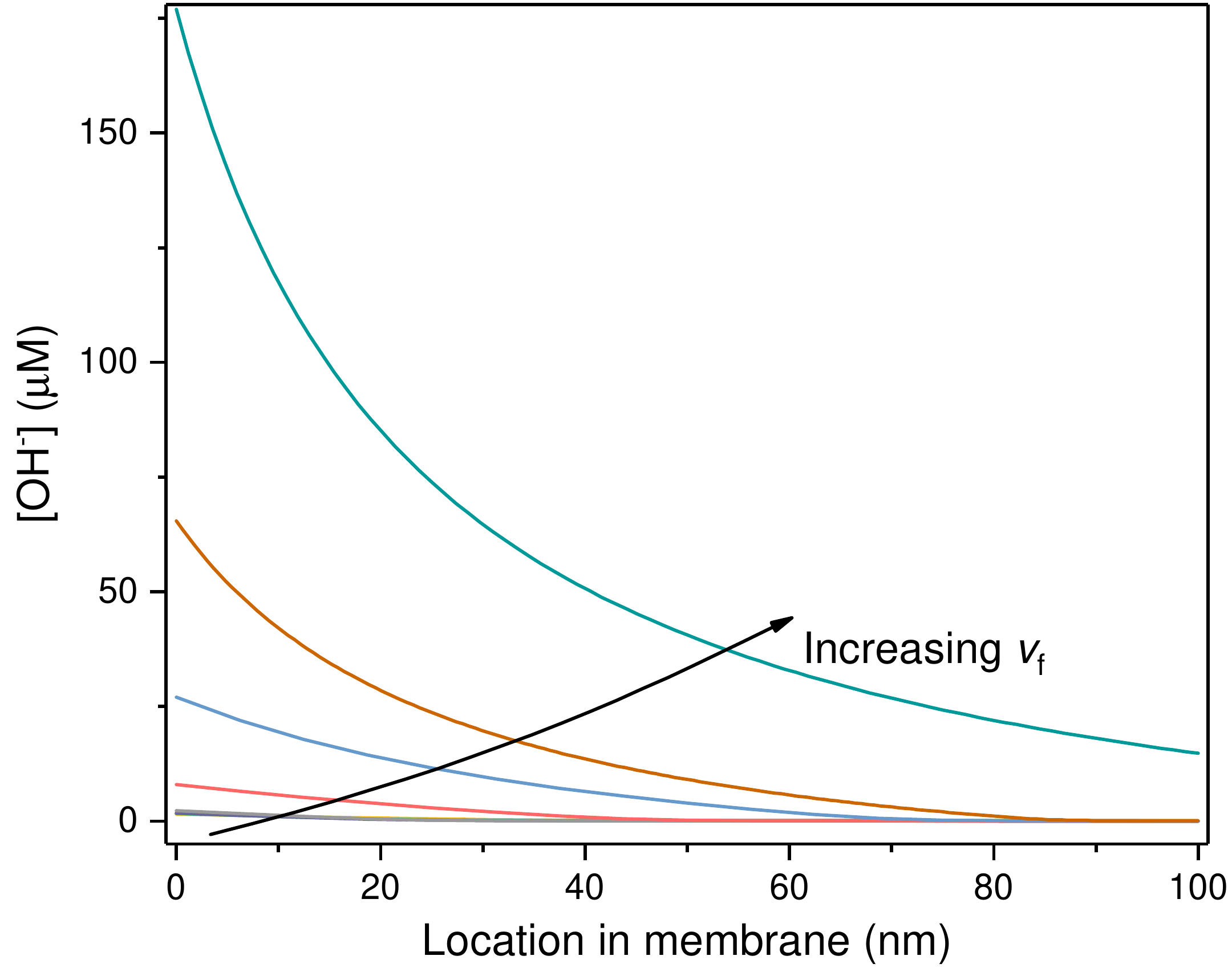}
	  \caption{\ce{OH-}}
	  \end{subfigure}
	  \begin{subfigure}{.5\textwidth}
  \centering
\includegraphics[width=1.0\linewidth]{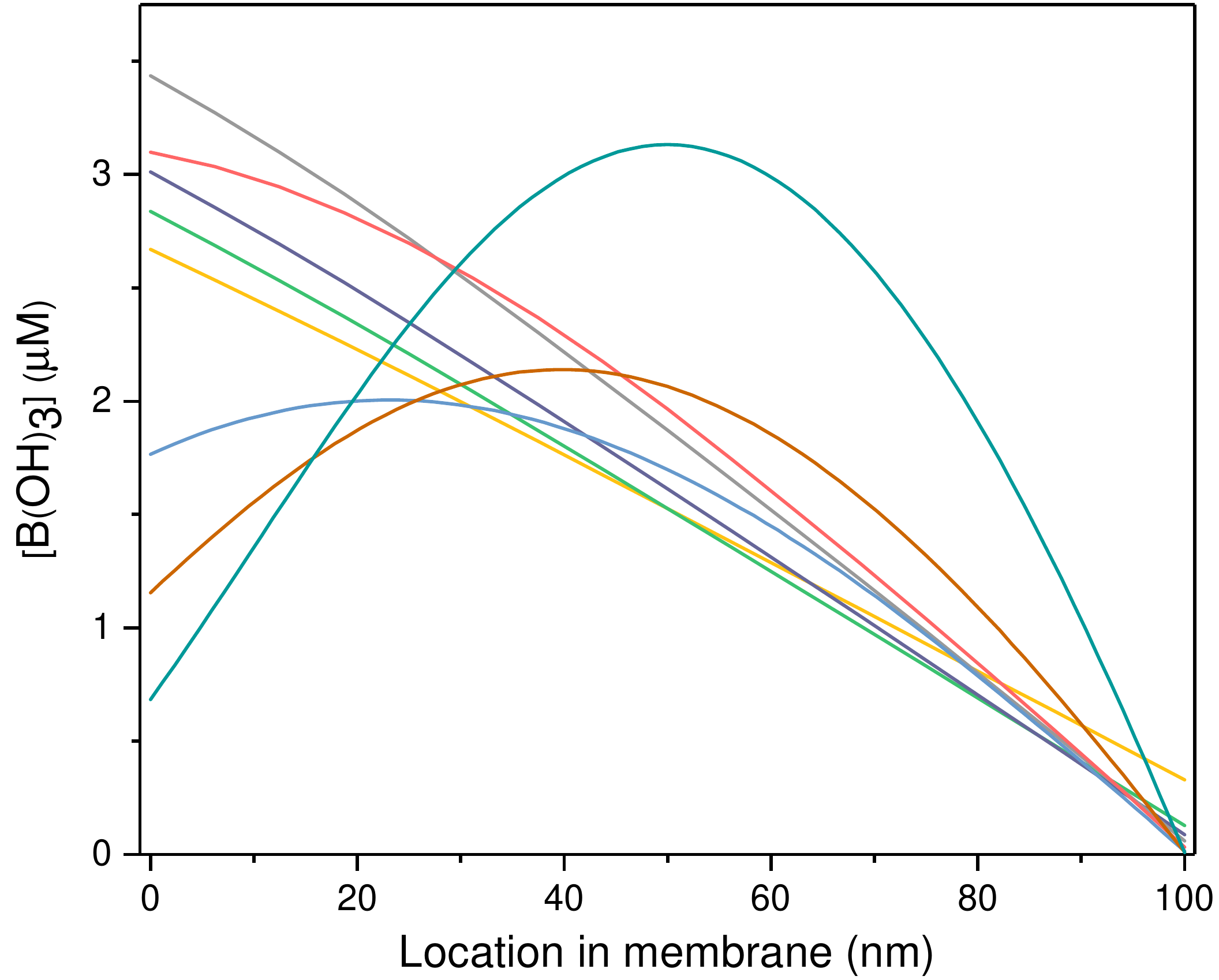}
	  \caption{\ce{B(OH)3}}
	  \end{subfigure}%
	  \begin{subfigure}{.5\textwidth}
  \centering
\includegraphics[width=1.0\linewidth]{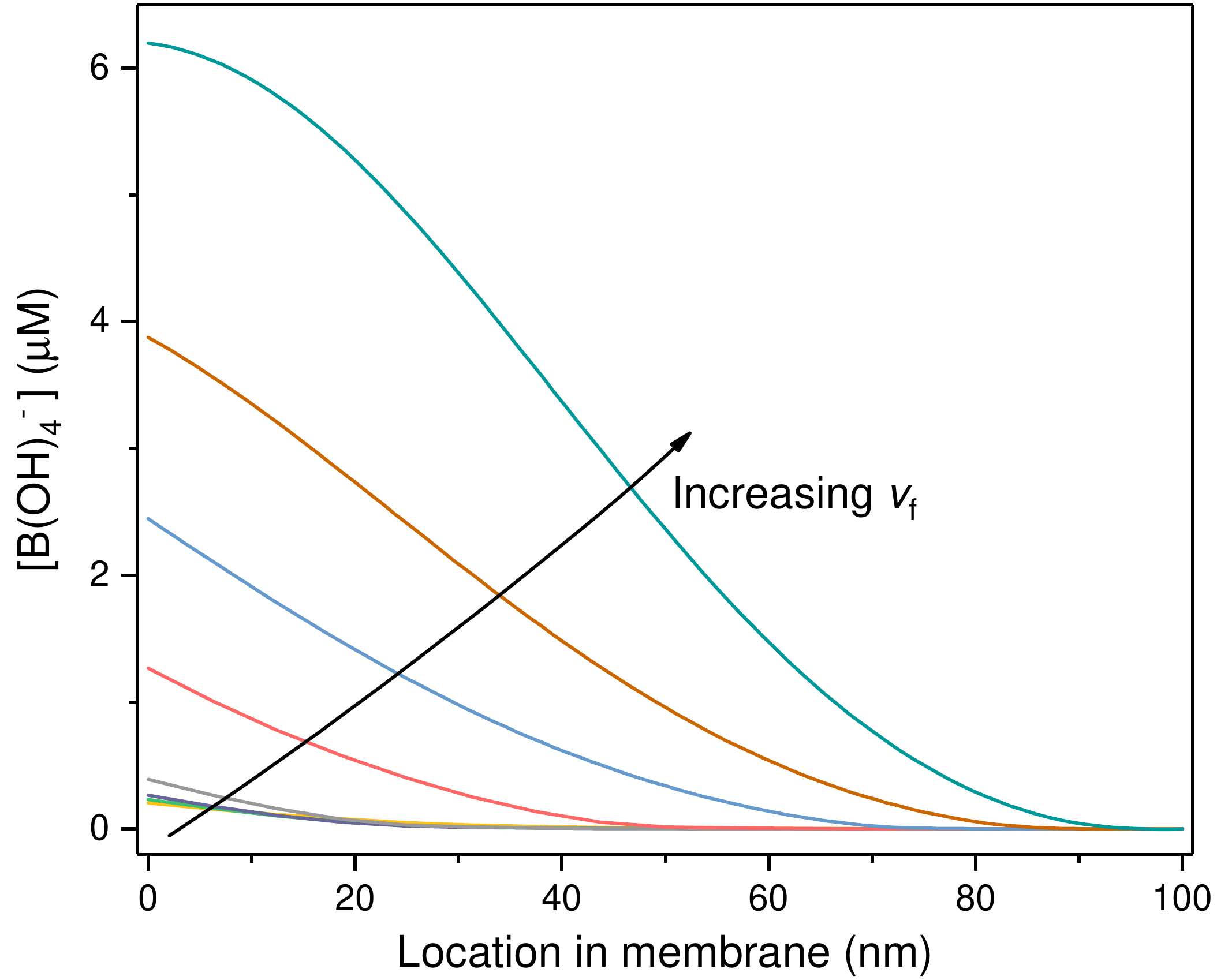}
	  \caption{\ce{B(OH)4-}}
	  \end{subfigure}
	  
\caption{Concentration profiles in the membrane of \ce{Na+}, \ce{Cl-}, \ce{H3O+}, \ce{OH-}, \ce{B(OH)3} and \ce{B(OH)4-} for different values of the water flow rate $v\s{f}$.}
\label{fig:ci1}
\end{figure}

\begin{figure}[H]
  \centering
  
  \begin{subfigure}{.33\textwidth}
  \centering
\includegraphics[width=1.0\linewidth]{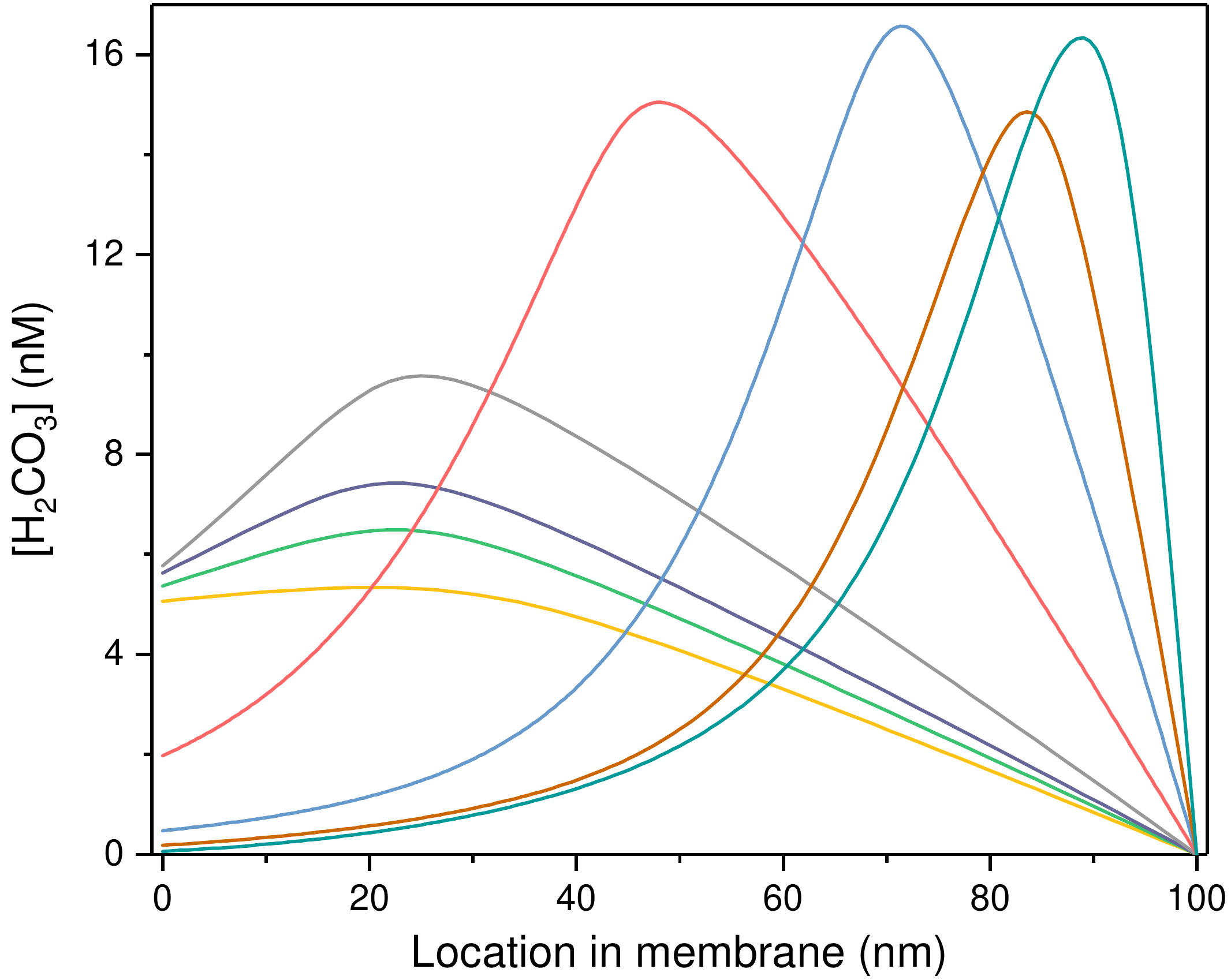}
	  \caption{\ce{H2CO3}}
	  \end{subfigure}%
	  \begin{subfigure}{.33\textwidth}
  \centering
\includegraphics[width=1.0\linewidth]{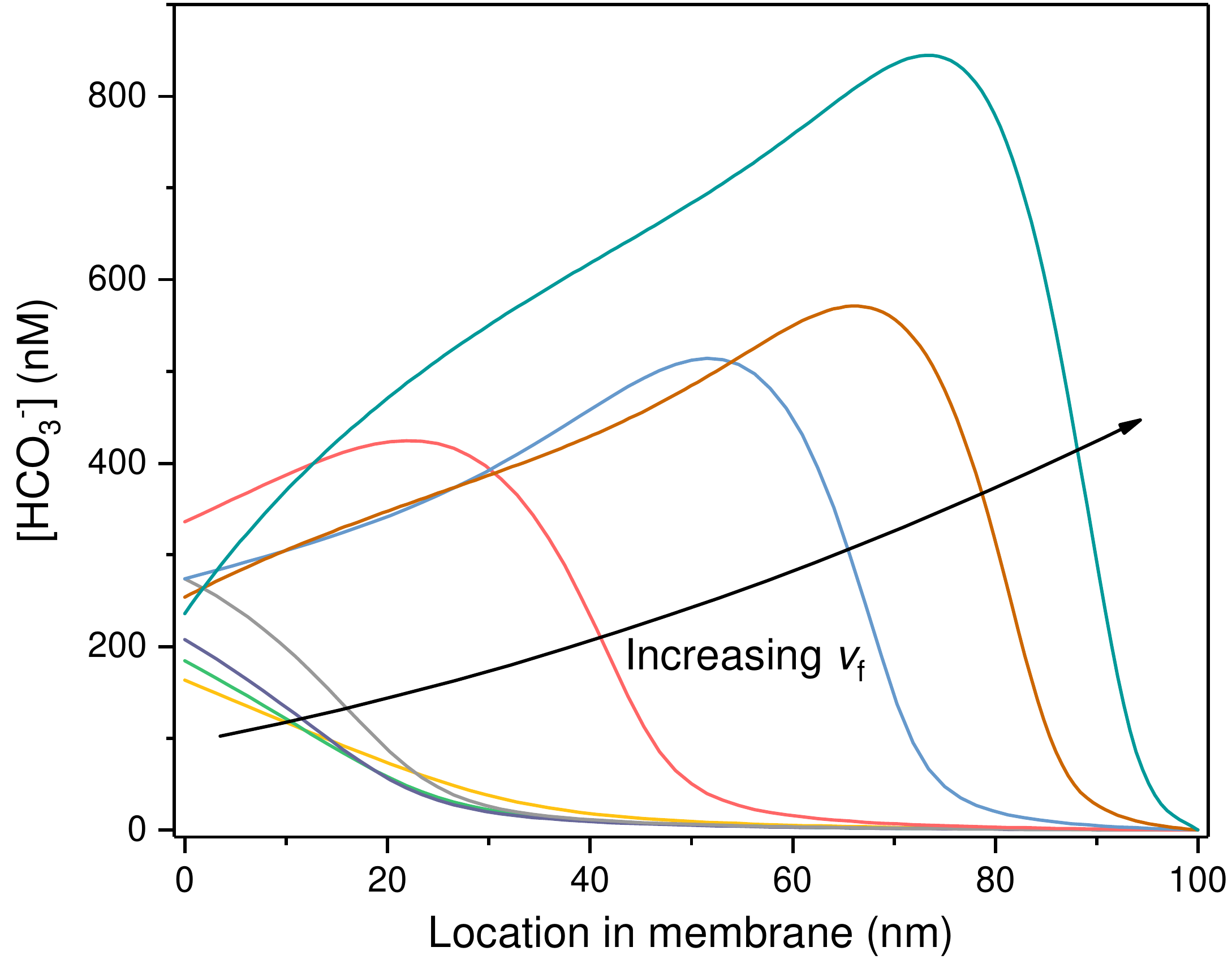}
	  \caption{\ce{HCO3-}}
	  \end{subfigure}%
	  \begin{subfigure}{.33\textwidth}
  \centering
\includegraphics[width=1.0\linewidth]{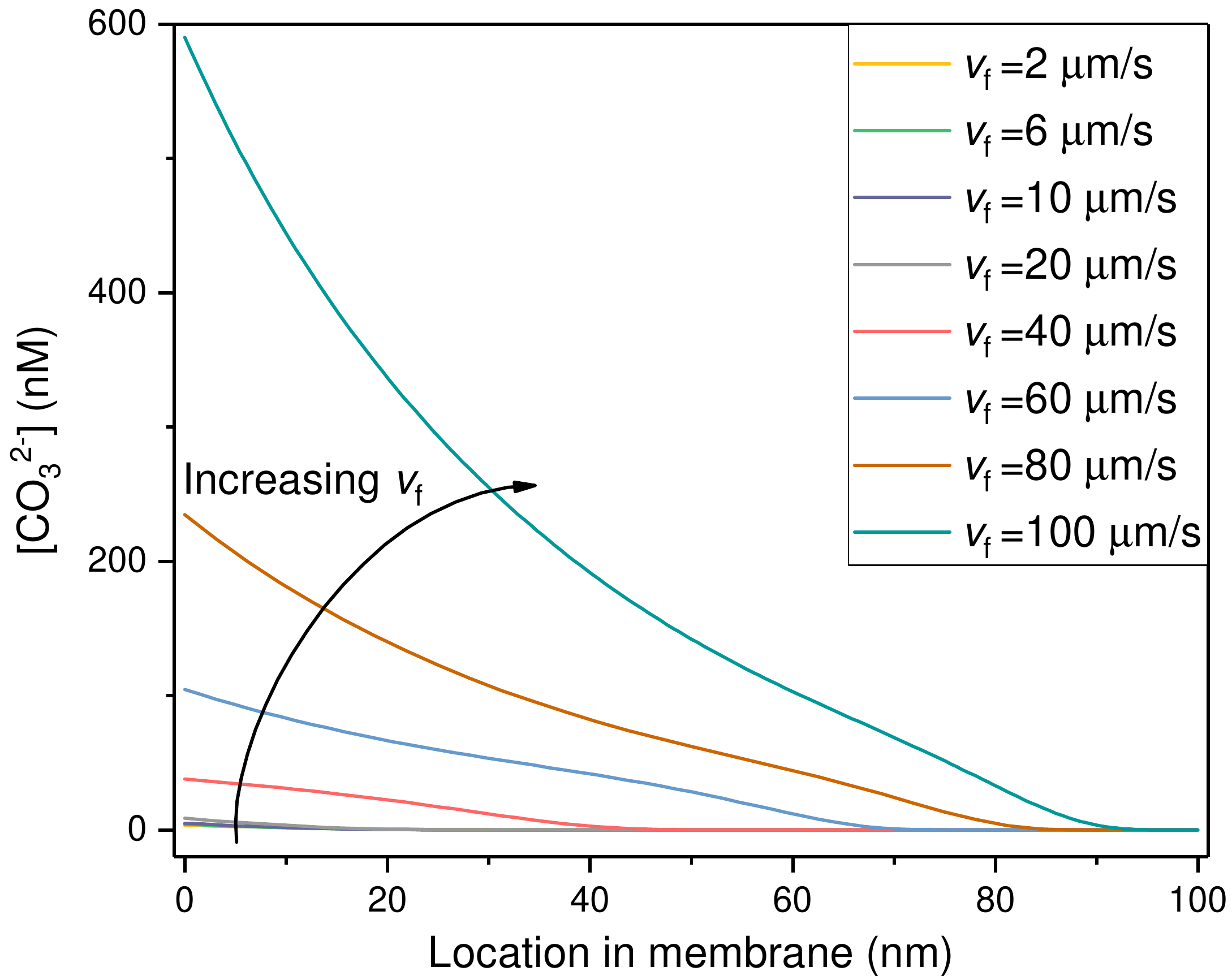}
	  \caption{\ce{CO3^2-}}
	  \end{subfigure}
\caption{Concentration profiles in the membrane of \ce{H2CO3}, \ce{HCO3-} and \ce{CO3^2-} for different values of the water flow rate $v\s{f}$.}
\label{fig:ci2}
\end{figure}

\begin{figure}[H]
\centering
\includegraphics[width=.7\linewidth]{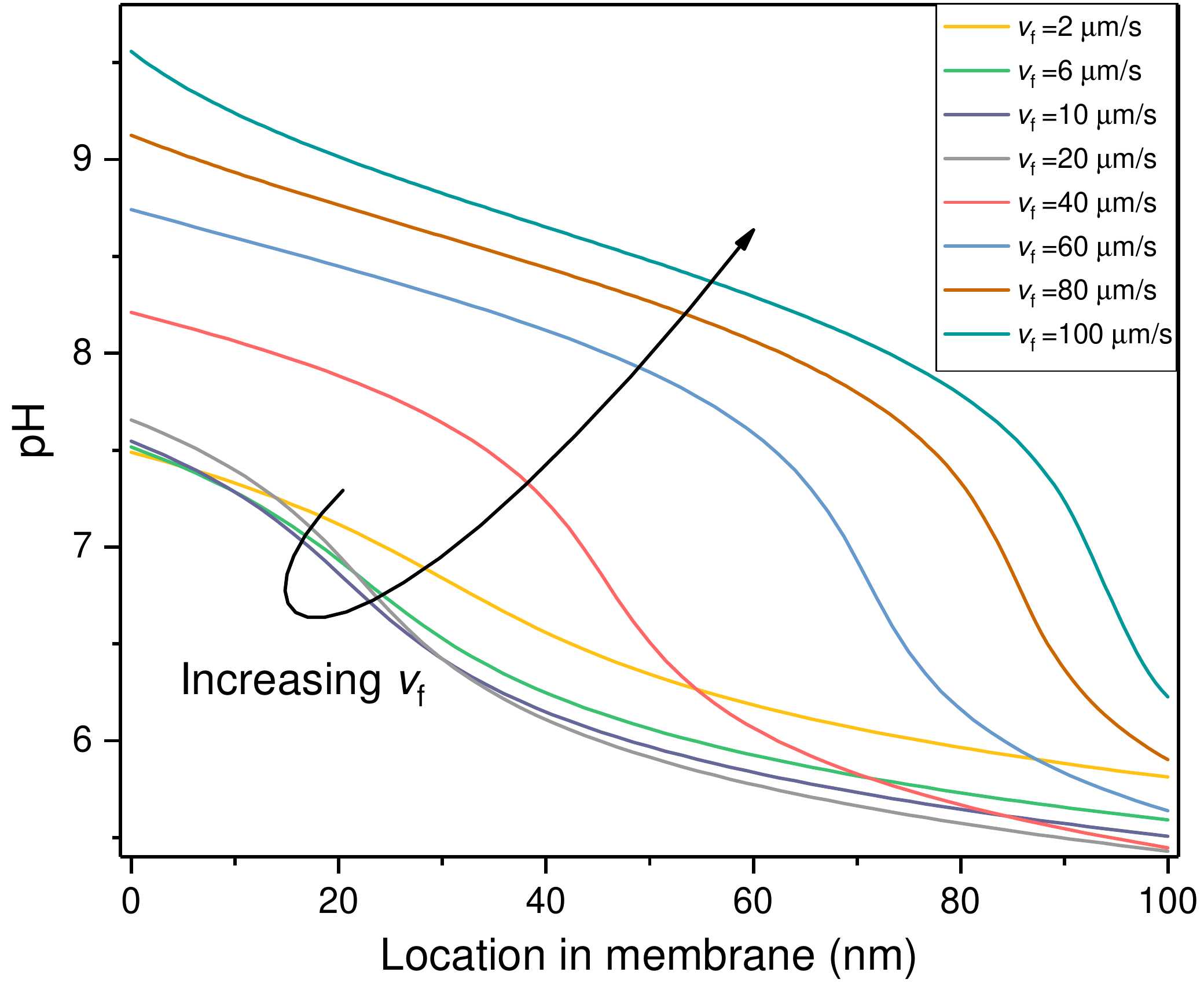}
\caption{\ce{H3O+} concentration in the membrane, expressed as pH, for different values of the water flow rate $v\s{f}.$} \label{fig:H}
\end{figure}

Figs. \ref{fig:ci1}-\ref{fig:H} show the obtained concentration profiles of the different ionic species in the membrane, for different water flow rates, $v\s{f}$. Such a representation of calculation output was not found in surveyed literature and highlights the complex character of multi-component transport in membrane processes. In general, it can be seen that when $v\s{f}$ increases so does the ion concentration at the beginning of the membrane, except for the hydronium ion. %

When we look at two neutral compounds, boric acid (Fig. \ref{fig:ci1}) and carbonic acid (Fig. \ref{fig:ci2}), we see that as $v\s{f}$ increases so does the concentration at the beginning of the membrane, until the trend reverses for $v\s{f}>20~\mu$m/s. With higher $v\s{f}$, the concentration profile for boric acid becomes parabolic-like. For carbonic acid and bicarbonate, their concentration profiles become non-monotonic and a maximum concentration is found in the membrane, the location of which is being displaced deeper into the membrane layer as the water flow rate increases.

At low water flow rates ($v\s{f}\lesssim 36~\mu$m/s) pH at the beginning of the membrane is lower than pH\textsubscript{feed}. For higher water flow rates the solution becomes more basic at the beginning of the membrane, but always remains more acidic than the feed when at the other side of the membrane. At the same time, pH\textsubscript{permeate} was always higher than in the feed.%

\subsection{Sensitivity analysis}

\begin{figure}[H]
\centering
\includegraphics[width=1\linewidth]{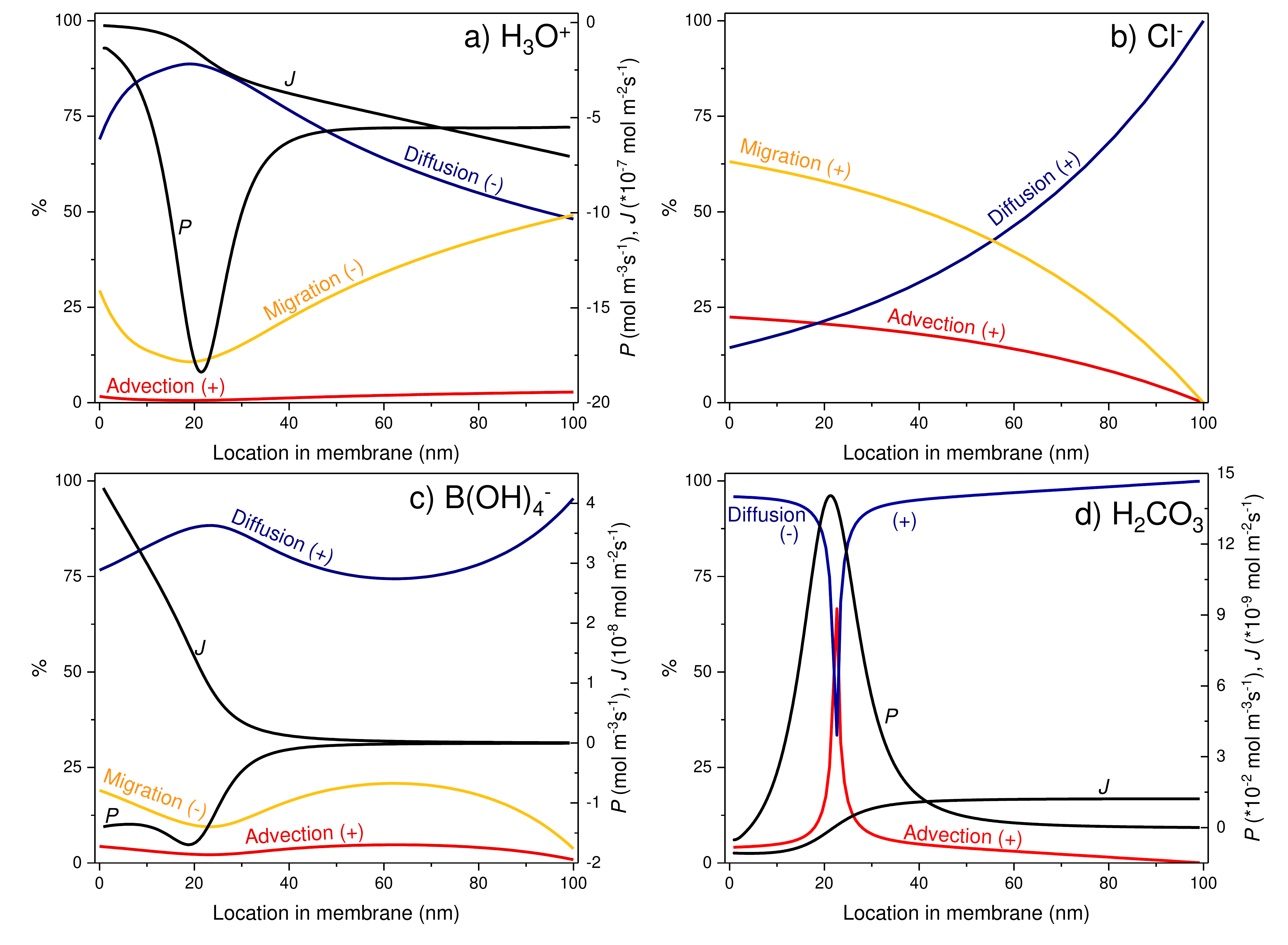}
	  
\caption{Relative contribution of advection, diffusion and migration to the flux of selected ions \ce{H3O+}, \ce{Cl-}, \ce{B(OH)4-} and \ce{H2CO3}. Signs next to each line describe whether the term helps to move the ion towards the permeate (+) or towards the feed (-).}
\label{fig:Sensitivity1}
\end{figure}

An interesting question is, what is the relative contribution of the three mechanisms that transport ions (advection, diffusion and migration) through the membrane. To analyze that question, the \textit{magnitudes of} the three terms in Eq.~\eqref{eq:v_i} are separately calculated and compared. Results in Fig. \ref{fig:Sensitivity1} are based on the standard condition described in Table~\ref{Inputs} with water flow rate $v\s{f}=10~\mu$m/s. We present here only diagrams for selected species. For the other species, the behavior is as follows. For \ce{Na+}, the contribution of each mechanism is invariant across the membrane with 50\% migration, 27\% advection, and 23\% diffusion; for \ce{OH-}, diffusion and migration are approx. the same as for \ce{H3O+} and advection is slightly higher; profiles for \ce{HCO3-} and \ce{CO3^2-} are similar to \ce{B(OH)4-}; for \ce{B(OH)3} there is no migration, and diffusion linearly increases from 90\% at the entrance to 100\% at the exit of the membrane.

As Fig. \ref{fig:Sensitivity1} shows, the transport of all species that are present at low concentration is dominated by diffusion. This contribution of diffusion was also discussed for \ce{H3O+} and \ce{OH-} in ref.~\cite{Nir2015b}. Other than that, the relative contribution of each term is not directly intuitive, for instance for \ce{Cl-} changing strongly across the membrane, while for carbonic acid, an extremely sharp change is observed due to the fact that at that point the diffusional contribution is zero, and only advection plays a role.  Clearly the study of the relative importance of the various mechanisms driving an ion, is non-trivial and may lead to interesting insights. Furthermore, three of the four ions discussed in Fig.~\ref{fig:Sensitivity1} are part of an acid/base equilibrium, and thus they can react away or be formed within the toplayer. This implies their flux, $J$, changes with position, as shown in Fig.~\ref{fig:Sensitivity1} because of a chemical production term, $P$. Because we consider steady-state, the chemical production term $P$ is equal to the gradient in $J$.

\subsection{Effect of membrane charge}

As TFC membranes are charged, we decided to examine the behavior and performance if the membrane charge would change. The standard values of the three functional groups' concentrations, $X\s{k}$, were multiplied by a certain factor $\alpha$, thus keeping their relative concentrations the same. We recorded the rejection of boron (Fig. \ref{fig:RBRNa_combined}B) and interestingly, for the current value of the membrane charge, boron rejection is almost at a maximum,  which we locate at $\alpha = 1.5$ to be $R\s{B}=93.6\%$. Even reducing the membrane charge dramatically, there is not much of a reduction in the ability of the membrane to reject boron with $R\s{B}=92\%$ at a 100 times reduced membrane charge. This can be explained by the fact that the neutral boron in the form of boric acid passes the membrane while not being affected by charge. For all values of $\alpha$ considered, salt rejection  remains very high, at values above 99.5\%. %

Calculation results for different values of pH of seawater (keeping the total carbonate concentration the same) are that for a 50$\times$ reduced membrane charge, $R\s{B}$ is not affected much by pH$\s{feed}$ in the entire range studied of 4<pH$\s{feed}$<10 [not shown]. However, for the standard value of membrane charge, and one with 10$\times$ more charge, for pH$\s{feed}$ less than 8.0, $R\s{B}$ was not changed, but for pH$\s{feed}$ beyond 8, boron rejection went up significantly, up to 99\% at pH$\s{feed}=10$. This latter result may not be of much practical relevance, because increasing the pH of seawater would lead to severe scaling of the system. % 

\begin{figure}[H]
  \centering
\includegraphics[width=0.6\textwidth]{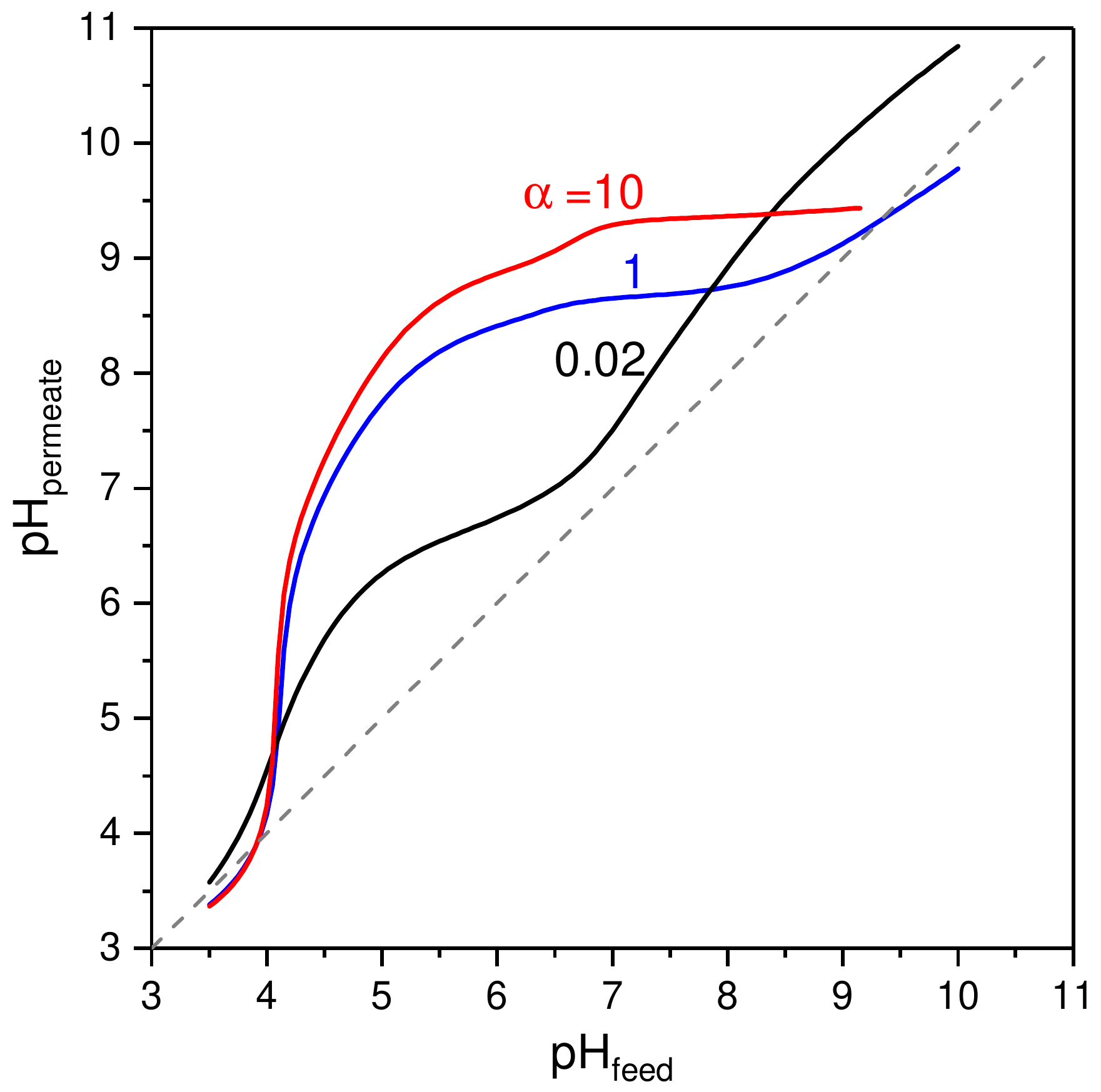}
	  \caption{pH\textsubscript{permeate} as a function of pH\textsubscript{feed}, for three different values of the relative membrane charge $\alpha$, for conditions of Table~\ref{Inputs}, $v\s{f}=10~\mu$m/s.} 	  \label{fig:pHpvspHf}
\end{figure}

Finally, we present our results for the effect of pH\textsubscript{feed} on pH\textsubscript{permeate} for three different values of the membrane charge (Fig. \ref{fig:pHpvspHf}). Except for very extreme pH$\s{feed}$, in all cases pH\textsubscript{permeate} is higher than pH\textsubscript{feed}. For pH\textsubscript{feed}$\sim$8.0, the increase is around one pH-point, which drops further at higher pH\textsubscript{feed}. For lower pH\textsubscript{feed}, down to pH 4, for the membranes with the original charge or larger, the permeate pH can be higher by around 3 pH points, while for the (almost) uncharged membrane, effluent pH is not more than 1 pH point higher than that of the feed. % 
Our calculation results can also be compared to two literature sources for experimental data. For the standard value of the membrane charge ($\alpha = 1$), measurements 
for SWRO are well reproduced by our calculations: for pH\textsubscript{feed}$=8.0$ we predict pH$\s{permeate}=8.7$, while experimentally a value of pH$\s{permeate}=8.6$--8.8 is reported \cite{Andrews2008}, and for pH$\s{feed}=9.0$, we predict pH$\s{permeate}=9.1$ while the experimental value is pH$\s{permeate}=9.1$--9.5~\cite{Nir2014a}. 

\section{Conclusions}

In this work we have developed and applied a theoretical transport model to describe the desalination of seawater using a flat sheet TFC-SWRO membrane. The water matrix was chosen to mimic seawater composition and contains high concentrations of NaCl, which is contrary to most other studies, which model salt rejection in more dilute systems. We investigated the influence of water flowrate (pressure), membrane charge and seawater pH. The water equilibrium and two weak acid systems - boric and carbonic acid - were considered additionally to the major ionic constituents. In the model, the membrane was considered to be a tortuous-porous polymeric structure, that holds a fixed amount of pH-dependent charge. Transport in the membrane was described by a Maxwell-Stefan approach including three driving forces contributing to transport: advection, diffusion and electromigration. Molecules were considered to travel through the membrane while retaining their full hydration shell, except for carbonate and bicarbonate for which we had to assume a reduction in the hydrated size to make them fit into the pores. Species entry in the membrane was corrected for by considering the partitioning factor based on the size of ions. Hindered transport of ions in membrane pores was accounted for by hydrodynamic correction factors for diffusion and convection. Electrical potential differences across the membrane-solution interfaces, caused by the Donnan effect, were taken into consideration as well. 

The model shows that in general diffusion is the dominant driving force of transport of ions in the membrane, while as the velocity of the fluid increased, ion concentration profiles become steeper. pH was, for all fluid velocities, lower in the membrane than in the feed, but on the permeate side was higher than feed pH. An increase in the membrane charge does not improve boron rejection but a lower membrane charge elevates permeate pH which might be helpful if second stage RO stage is employed for boron removal. Neither changes in pH or in membrane charge affect the salt rejection significantly which always remained above 99\%. pH of the permeate was successfully predicted by the model, which can be helpful in the planning of post treatment stages such as pH stabilization.

In order to validate the model in more detail, experimental verification of observed relations is relevant. It is important to note that the model relies on physico-chemical properties such as chemical equilibrium constants and hydrated ion sizes, for which some of those parameters are more sensitive than others to temperature changes and so will have to be adjusted for systems that differ greatly from a temperature 25 $^\circ$C, which was the temperature considered in this work.

%\section*{Acknowledgments}

%\bibliographystyle{ieeetr}%abbr % reference list style
%\bibliography{Thesis} %makes the reference list

\end{document}